\def\norm[#1]{\lVert{#1} \rVert}

\documentclass[draftcls,onecolumn,11pt]{IEEEtran}
\usepackage{mathtools}
\usepackage{amsthm}

\usepackage{hyperref}

\usepackage{stfloats}
\usepackage{xcolor}
\usepackage{amsfonts}
\usepackage{acronym}
\usepackage{cite}
\usepackage{amsmath,amssymb}
\usepackage[dvips]{graphicx}
\usepackage[caption=false,font=footnotesize]{subfig}
\usepackage[geometry]{ifsym}
\usepackage{flushend}

\usepackage{bm,upgreek}
\makeatletter
\newcommand{\bfgreek}[1]{\bm{\@nameuse{up#1}}}
\makeatother
\DeclareMathSymbol{\Theta}{\mathalpha}{operators}{2}
\begin{document}

\title{Hybrid 3D Localization for Visible Light Communication Systems}

\author{\authorblockN{Alphan \c{S}ahin, Yusuf Said Ero\u{g}lu,  {\.I}smail G\"{u}ven\c{c}, Nezih Pala, and Murat Y\"{u}ksel\\}
\thanks{This project was supported by National Science Foundation awards CNS-1422062 and CNS-1422354, and 2014 Ralph E. Powe Junior Faculty Enhancement Award.}
\thanks{Alphan \c{S}ahin, Yusuf Said Ero\u{g}lu,  {\.I}smail G\"{u}ven\c{c}, and Nezih Pala are with the Department of Electrical and Computer Engineering, Florida International University, Miami, FL (email: { \{asahin,yerog001,iguvenc,npala\}@fiu.edu}).
}
\thanks{Murat Y\"{u}ksel is with the
Department of Computer Science and Engineering, University of Nevada, Reno, NV
(email: { yuksem@cse.unr.edu}).
}
}

\maketitle
\begin{abstract}
In this study, we investigate hybrid utilization of angle-of-arrival (AOA)  and received signal strength (RSS) information in visible light communication (VLC) systems for 3D localization. We show that AOA-based localization method allows the receiver to locate itself via a least squares estimator by exploiting the directionality of light-emitting diodes (LEDs). We then prove that when the RSS information is taken into account, the positioning accuracy of AOA-based localization can be improved further using a weighted least squares solution. On the other hand, when the radiation patterns of LEDs are explicitly considered in the estimation, RSS-based localization yields highly accurate results. In order to deal with the system of non-linear equations for RSS-based localization, we develop an analytical learning rule based on the Newton-Raphson method. The non-convex structure is addressed by initializing the learning rule based on 1) location estimates, and 2) a newly developed method, which we refer as  random report and cluster algorithm. As a benchmark, we also derive analytical expression of the Cram\'er-Rao lower bound (CRLB) for RSS-based localization, which captures any deployment scenario positioning in 3D geometry. Finally, we demonstrate the effectiveness of the proposed solutions for a wide range of LED characteristics and orientations through extensive computer simulations.

\end{abstract}

\begin{IEEEkeywords}
CRLB, estimation, free space optics (FSO), Lambertian pattern, localization, positioning, VLC.
\end{IEEEkeywords}

\acrodef{BLUE}{best linear unbiased estimator}
\acrodef{CRLB}{Cram\'er-Rao lower bound}
\acrodef{VLC}{visible light communication}
\acrodef{DoF}{degrees-of-freedom}
\acrodef{CSI}{channel state information}
\acrodef{PDP}{power delay profile}
\acrodef{CIR}{channel impulse response}
\acrodef{PAPR}{peak-to-average power ratio}
\acrodef{OOB}{out-of-band}
\acrodef{HCF}{half-cosine function}
\acrodef{VSB}{vestigial side-band modulation}
\acrodef{GFDM}{generalized frequency division multiplexing}
\acrodef{PSWF}{Prolate spheroidal wave function}
\acrodef{OFDM}{orthogonal frequency division multiplexing}
\acrodef{OFDMA}{orthogonal frequency division multiple accessing}
\acrodef{BFDM}{Bi-orthogonal frequency division multiplexing}
\acrodef{DFT}{discrete Fourier transformation}
\acrodef{IDFT}{inverse discrete Fourier transformation}
\acrodef{IFFT}{inverse fast Fourier transformation}
\acrodef{FBMC}{filter bank multicarrier}
\acrodef{CP}{cyclic prefix}
\acrodef{QAM}{quadrature amplitude modulation}
\acrodef{OQAM}{offset quadrature amplitude modulation}
\acrodef{FFT}{fast Fourier transformation}
\acrodef{CFO}{carrier frequency offset}
\acrodef{SIR}{signal-to-interference ratio}
\acrodef{SNR}{signal-to-noise ratio}
\acrodef{SINR}{signal-to-interference-plus-noise ratio}
\acrodef{ICI}{inter-carrier interference}
\acrodef{ISI}{inter-symbol interference}
\acrodef{PPN}{polyphase network}
\acrodef{WSSUS}{wide-sense stationary uncorrelated scattering}
\acrodef{SEM}{spectral emission mask}
\acrodef{BWA}{broadband wireless access}
\acrodef{PSD}{power spectral densitie}
\acrodef{MIMO}{multiple-input multiple-output}
\acrodef{DSL}{digital subscriber lines}
\acrodef{OFDP}{optimal finite duration pulse}
\acrodef{FMT}{Filtered multitone}
\acrodef{SMT}{Staggered multitone}
\acrodef{CMT}{Cosine-modulated multitone}
\acrodef{IOTA}{Isotropic orthogonal transform algorithm}
\acrodef{RMS}{root mean square}
\acrodef{MMSE}{minimum mean square error}
\acrodef{MLD}{maximum likelihood detection}
\acrodef{STBC}{space-time block coding}
\acrodef{TO}{timing offset}
\acrodef{CFO}{carrier frequency offset}
\acrodef{IAM}{interference approximation method}
\acrodef{PN}{phase noise}
\acrodef{RF}{radio frequency}
\acrodef{CPM}{continuous phase modulation}
\acrodef{VAP}{visible light access point}
\acrodef{ADC}{analog-to-digital converter}
\acrodef{PA}{power amplifier}
\acrodef{CCDF}{complementary cumulative distribution function}
\acrodef{SC-FDMA}{single carrier frequency division multiple accessing}
\acrodef{CP-OFDM}{Cyclic Prefix-OFDM}
\acrodef{ZP-OFDM}{Zero Padded-OFDM}
\acrodef{FPGA}{field-programmable gate array}
\acrodef{STTC}{space-time trellis coding}
\acrodef{BER}{bit error rate}
\acrodef{ZP}{zero-padded}
\acrodef{FDE}{frequency domain equalization}
\acrodef{TDE}{time domain equalization}
\acrodef{LS}{least squares}
\acrodef{SC-FDE}{single carrier frequency domain equalization}
\acrodef{FB-S-FBMC}{filter-bank-spread-filter-bank multicarrier}
\acrodef{LTE}{Long Term Evolution}
\acrodef{AWGN}{additive white Gaussian noise}
\acrodef{RD}{random demodulator}
\acrodef{FTN}{faster-than Nyquist}
\acrodef{PRS}{partial-response signaling}
\acrodef{MAP}{maximum a posteriori}
\acrodef{DFE}{decision feedback equalizer}
\acrodef{FIR}{finite impulse response}
\acrodef{LMS}{least-mean-square}
\acrodef{RLS}{recursive-least-squares}
\acrodef{PHYDYAS}{Physical layer for dynamic access}
\acrodef{MLSE}{maximum likelihood sequence estimator}
\acrodef{EGF}{Extended Gaussian function}
\acrodef{AIC}{Akaike information criterion}
\acrodef{SIC}{successive interference cancellation}
\acrodef{CDMA}{code division multiple accessing}
\acrodef{i.i.d.}{independent and identically distributed }
\acrodef{LED}{light emitting diode}
\acrodef{FOV}{field of view}
\acrodef{PD}{photo detector}
\acrodef{RSS}{received signal strength}
\acrodef{LOS}{line-of-sight}
\acrodef{MSE}[MSE]{mean square error}
\acrodef{RMSE}[RMSE]{root mean square error}
\acrodef{ML}[ML]{Maximum likelihood}
\acrodef{NLS}[NLS]{nonlinear least squares}
\acrodef{FIM}{Fisher Information Matrix}
\acrodef{GPS}{Global Positioning System}
\acrodef{UWB}{ultra wideband}
\acrodef{WLAN}{wireless local area network}
\acrodef{RFID}{radio frequency identification}
\acrodef{CDF}{cumulative distribution function}
\acrodef{AOA}{angle of arrival}
\acrodef{RRC}{random report and cluster}
\acrodef{PMF}{probability mass function}
\acrodef{PDF}{probability density function}

\def\numberOfTransmitters{K}
\def\numberOfLEDs{M}
\def\indexLED{m}
\def\indexTransmitter{k}

\def\locationLED[#1]{{\rm \bf r}_{#1}}
\def\locationDetector{{\rm \bf r}_{\rm R}}
\def\directionLED[#1]{{\rm \bf n}_{#1}}
\def\directionDetector{{\rm \bf n}_{\rm R}}
\def\incidenceVector[#1]{{\rm \bf v}_{#1}}

\def\directionLEDElements[#1][#2]{{n}_{{#1}}^{(#2)}}
\def\directionDetectorElements[#1]{{n}_{{\rm R}}^{(#1)}}
\def\incidenceElements[#1][#2]{{#2}_{#1}}

\def\loglikelihood[#1]{\mathcal{L}\left(#1\right)}

\def\distance[#1]{R_{#1}}
\def\Ar{A_{\rm R}}
\def\FOV{\theta_{\rm FOV}}
\def\mode{n}

\def\locationDetectorEstimate{\tilde{{\rm \bf r}}_{\rm R}}
\def\polarAngle{\theta_{\rm polar}}
\def\ceilingAngle{\theta_{\rm ceiling}}

\def\power[#1]{{P_{#1}}}
\def\angleLOSandTX[#1]{\phi_{#1}}
\def\angleLOSandRX[#1]{\theta_{#1}}

\def\rect[#1]{{\Pi}\left(#1\right)}

\def\complexNumbers{\mathbb{C}}
\def\realNumbers{\mathbb{R}}

\def\function[#1]{f\left(#1\right)}

\def\parameterVector{{ \bfgreek{theta}}}
\def\observationVector{{\rm \bf s}}
\def\powerMatrix{{\rm \bf P(\parameterVector)}}
\def\powerVector[#1]{{{\rm \bf p}(#1)}}
\def\noiseVector{{\rm \bf n}}
\def\identityMatrix[#1]{{\rm \bf I}_{#1}}
\def\noiseVariance{\sigma_{\rm n}^2}
\def\realGaussian[#1][#2]{\mathcal{N}\left(#1,#2\right)}
\def\probability[#1]{p\left(#1\right)}
\def\vectorization[#1]{{\rm vec}\left(#1\right)}

\def\kernel[#1]{\ker\left(#1\right)}
\def\covarianceMatrix{{\rm \bf C}}

\def\anObservationVector{{\rm \bf x}}
\def\meanFunctionOfParameters{{\boldsymbol \mu}(\parameterVector)}
\def\coVarianceFunctionOfParameters{{\rm \bf C}(\parameterVector)}
\def\FisherInformationMatrix{{\rm \bf J({\parameterVector})}}
\def\inverseFisherInformationMatrix{{\rm \bf J}^{-1}({\parameterVector})}
\def\jacobianMatrix{{\rm \bf H}}

\def\expectedOperator[#1]{
\mathbb{E}\left[{#1}\right]
}

\def\lineLED[#1]{L_{#1}}
\def\projectionlineLEDNull[#1]{{\rm \bf A}_{#1}}
\def\distanceNull[#1]{d_{#1}}
\def\distanceVector[#1]{{\rm \bf d}_{#1}}
\def\intersectionPoint[#1]{{\rm \bf b}_{#1}}
\def\setSelectLED[#1]{s_{#1}}
\def\derivativeFunction[#1][#2][#3][#4][#5][#6][#7]{g\left(#1;#2,#3,#4,#5,#6,#7\right)}

\section{Introduction}
The reliable positioning in the environments where the \ac{GPS} signals cannot easily penetrate has become a growing research area due to its numerous applications \cite{Liu_Survey2007}.  Among many other technologies, \ac{RFID}\cite{Hahnel_2004,Bouet_2008,Bekir_WCNC_2015}, \ac{UWB} positioning \cite{Gezici_2005}, cellular or \ac{WLAN} based positioning \cite{Sayed_2005, guvenc2009survey}, and fingerprint based localization techniques \cite{Kaemarungsi_2005} are the most prominent ones for \ac{GPS}-free localization. Beside those technologies, \ac{VLC} \cite{Komine_2004,2013-sevincer-lightnets, Burchardt_2014}, which has been recently emerging as a promising technology to provide high data rates for short range communications, can also deliver high accuracy for 3D localization \cite{Elgala_2011, liu2014cellular}. 

\ac{VLC} systems have numerous advantages in comparison to other legacy systems. First, the performance of \ac{VLC} systems is not affected by the \ac{RF} interference. Moreover, multipath fading can be averaged out as the area of photodetectors are very large compared to wavelength of the visible light \cite{kahn_1997}. Second, \acp{LED} offer narrow beamwidth, which enables more accurate angle of arrival information at the receiver side \cite{eroglu_2015,Bilgi_2012}. Third, \ac{VLC} systems can be utilized in scenarios where \ac{RF} radiation is potentially hazardous or even forbidden such as planes or hospitals. Lastly, \ac{VLC} systems maintain maintains high data rate communications while simultaneously providing illumination.

Due to its aforementioned advantages, employing \ac{VLC} systems for the localization is recently getting significant attention in the literature. In particular, two approaches have become prominent for \ac{VLC}-based localization: \ac{AOA} based methods \cite{Bilgi_2012,eroglu_2015}  and \ac{RSS} based methods \cite{zhou_2012, Zhang_2014}.
While \ac{AOA}-based approach takes the direction information of \ac{LED} transmitters, \ac{RSS}-based localization considers the captured power from \ac{LED} transmitters. For example, \color{black} 
in \cite{eroglu_2015}, \ac{AOA}-based localization is investigated for indoor scenarios and a connectivity-based solution is proposed under 2D settings. The proposed solution exploits the narrow \ac{FOV}s of \acp{LED} and relies on that fact that the receiver is likely to be located at the intersection of the directions of the connected LEDs transmitters. Yet, the proposed method is valid only under 2D settings where two straight lines always intersect between each other unless they are not parallel to each other. \color{black} 
In \cite{Bilgi_2012}, \ac{AOA} information of \ac{LED} transmitters are exploited in an ad-hoc networking environment. By starting from a limited number \ac{GPS}-enabled devices, each node discovers its own location and broadcasts this information in order to help the other nodes to find their own locations. 
In \cite{zhou_2012}, an \ac{RSS}-based indoor positioning algorithm based on preinstalled \ac{LED} transmitters is proposed to estimate receiver locations by analytically solving the equations that characterize the Lambertian pattern \cite{Barry_1993}. However, the analytical derivations are not generic, and cover only a limited number of scenarios, where the directionality of the \acp{LED} (i.e., their modes) and the receiver's direction are fixed to some predetermined values. In \cite{Zhang_2014}, \ac{CRLB} for \ac{RSS}-based localization is derived for a specific scenario of \ac{LED} transmitters and receiver positions, where the authors emphasize the complexity of obtaining the \ac{CRLB} in a 3D geometry. 

In this study, we evaluate \ac{AOA}-based and \ac{RSS}-based localization methods by using multi-element \acp{VAP} which consist of several \ac{LED} transmitters \cite{2013-sevincer-lightnets, Bilgi2014150, Tsonev_2013}. \color{black}The
main contributions of this paper are as follows: 
\begin{itemize}
    \item Hybrid Localization: We develop hybrid localization methods that utilize both \ac{AOA} and \ac{RSS} information in the estimation process to improve the positioning  accuracy in \ac{VLC} systems. For \ac{AOA}-based localization, we show that the \ac{VLC} receiver can estimate its own location via a \ac{LS} estimator and the  positioning accuracy can be improved further by incorporating with the \ac{RSS} information. In addition, we demonstrate that the outcome of \ac{AOA}-based location estimation can effectively be useful to deal with the non-convex objective function in \ac{RSS}-based localization.
	\item Theoretical Framework: We introduce a comprehensive theoretical framework that allows us to investigate the various hybrid localization methods for VLC and the impact of the orientations of LED transmitters and the physical characteristics of LEDs on the estimator performance. As opposed to earlier work in the literature, the proposed methods and theoretical investigations are generic, and applicable to \emph{any} 3D topology. Based on the introduced framework, we discuss the optimal LED weighting problem for \ac{AOA}-based localization and the non-convex objective function for \ac{RSS}-based localization. As a benchmark, we derive the analytical expression of \ac{CRLB} for \ac{RSS}-based localization, which generalizes the derivations provided in \cite{Zhang_2014}. By using the derived \ac{CRLB}, we demonstrate the trade-offs between light distribution and localization accuracy distribution and discuss the impact of \ac{LED} characteristics, such as directivity, density, and orientation, on the maximum achievable positioning accuracy. 
	\item Learning Algorithms: We introduce two algorithms, i.e., analytical learning algorithm based on Newton-Raphson method and \ac{RRC} algorithm, to combat with the system of non-linear equations and the non-convex structure of the objective function for RSS-based localization, respectively. In order to increase the likelihood that the Newton-Raphson method converges to the global optimum, we employ \ac{RRC} algorithm and show its effectiveness numerically. 
\end{itemize} \color{black}

The rest of the paper is organized as follows. The system model which captures  any \ac{VAP} deployment scenario in 3D geometry is provided in Section~\ref{sec:systemModel}. \ac{AOA}-based localization and hybrid utilization of \ac{AOA} and \ac{RSS} information for \ac{VLC} systems are discussed in Section \ref{sec:AOA}. Subsequently, \ac{RSS}-based localization is investigated in Section \ref{sec:RSSI}. Theoretical expression of \ac{CRLB} for \ac{RSS}-based localization, the analytical learning rule, and the \ac{RRC} algorithm are provided in this section. The numerical results evaluating the performance of the proposed methods for different configuration of \acp{VAP} are given in Section \ref{sec:numericalResults}. Finally,  some concluding remarks are indicated in Section \ref{sec:conclusions}.

Notations: $\identityMatrix[K]$ is the $K\times K$ identity matrix. The transpose operation is denoted by $(\cdot)^{\rm T}$. The Moore-Penrose pseudoinverse operation is denoted by $(\cdot)^{\rm \dagger}$. The operation of $\norm[\cdot]_2$ is the 2-norm of its argument. The vectorization and the expectation operators are denoted by  $\vectorization[\cdot]$ and $\expectedOperator[\cdot]$, respectively. The kernel of a matrix is represented by $\kernel[{\cdot}]$. The field of real numbers is shown as $\realNumbers$.
$\realGaussian[0][\covarianceMatrix]$ is the Gaussian distribution with zero mean and covariance matrix of $\covarianceMatrix$.

\section{System Model}
\label{sec:systemModel}

\begin{figure}[!t]
\centering
{\includegraphics[width =3.5in]{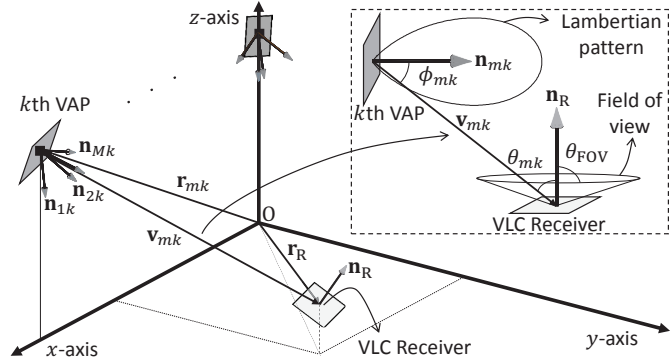}
}
\caption{System model with multiple VAPs and a single VLC receiver.}
\label{fig:systemModel}
\end{figure} 

We consider $\numberOfTransmitters$ \acp{VAP} communicating with a \ac{VLC} receiver as illustrated in \figurename~\ref{fig:systemModel}. Without loss of generality, the location of the receiver and its orientation are denoted by $\locationDetector=[x,y,z]^{\rm T}\in\realNumbers^{3\times1}$ and $\directionDetector=[\directionDetectorElements[x],\directionDetectorElements[y], \directionDetectorElements[z]]^{\rm T}\in\realNumbers^{3\times1}$ in Cartesian coordinate system, respectively. We consider multi-element \acp{VAP} which consist of $\numberOfLEDs$ \ac{LED} transmitters. The location of $\indexLED$th \ac{LED} transmitter of $\indexTransmitter$th \ac{VAP} and its  orientation  are denoted by $\locationLED[{\indexLED\indexTransmitter}]=[x_{\indexLED\indexTransmitter}, y_{\indexLED\indexTransmitter}, z_{\indexLED\indexTransmitter}]^{\rm T}\in\realNumbers^{3\times1}$ and $\directionLED[{\indexLED\indexTransmitter}]=[\directionLEDElements[\indexLED\indexTransmitter][x], \directionLEDElements[\indexLED\indexTransmitter][y], \directionLEDElements[\indexLED\indexTransmitter][z]]^{\rm T}\in\realNumbers^{3\times1}$, respectively.  By assuming transmit power of $1$~W for each \ac{LED} transmitter, the signal power of the $\indexLED$th LED of  $\indexTransmitter$th \ac{VAP} at the receiver is given by \cite{Barry_1993} 
\begin{align}
\power[\indexLED\indexTransmitter]=\frac{\mode+1}{2\pi}&\cos^\mode(\angleLOSandTX[\indexLED\indexTransmitter]) \cos(\angleLOSandRX[\indexLED\indexTransmitter])\frac{\Ar}{\distance[\indexLED\indexTransmitter]^2} \nonumber \\&~~~~~~~~~~~~~~~~\times\rect[{\frac{\angleLOSandRX[\indexLED\indexTransmitter]}{\FOV}}]
\rect[{\frac{\angleLOSandTX[\indexLED\indexTransmitter]}{\pi/2}}]
~,
\label{eq:power}
\end{align}
where $\angleLOSandTX[\indexLED\indexTransmitter]$ is the angle between the \ac{LED} transmitter orientation vector and the incidence vector, $\angleLOSandRX[\indexLED\indexTransmitter]$ is the angle between receiver orientation vector and  the incidence vector, $\distance[\indexLED\indexTransmitter]$ is the distance between the \ac{LED} transmitter and the receiver, $\Ar$ is the area of \ac{PD} in ${\rm m}^2$, $\FOV$ is the \ac{FOV} of \ac{PD}, $\mode$ is the mode number which specificities the directionality of \ac{LED}, and $\rect[\cdot]$ is the rectangle function defined as
\begin{align}
\rect[x]\triangleq\begin{cases}
1 & \mbox{for  } |x| \le 1 \\
0 & \mbox{for  } |x| > 1
\end{cases}~.
\end{align}
While $\rect[{{\angleLOSandRX[\indexLED\indexTransmitter]}/{\FOV}}]$ in \eqref{eq:power} implies that a \ac{VLC} receiver can detect the light only when $\angleLOSandRX[\indexLED\indexTransmitter]$ is less than $\FOV$,  $\rect[{{\angleLOSandTX[\indexLED\indexTransmitter]}/{(\pi/2)}}]$ ensures that the location of \ac{VLC} receiver is in the \ac{FOV} of \ac{LED} transmitter. Let $\incidenceVector[\indexLED\indexTransmitter]=\locationDetector- \locationLED[{\indexLED\indexTransmitter}]=[\incidenceElements[\indexLED\indexTransmitter][a], \incidenceElements[\indexLED\indexTransmitter][b], \incidenceElements[\indexLED\indexTransmitter][c]]^{\rm T}\in\realNumbers^{3\times1}$ be the incidence vector between the receiver and the $\indexLED$th \ac{LED} transmitter of $\indexTransmitter$th \ac{VAP}. Then, the parameters of \eqref{eq:power} can be expressed as 
\begin{align}
\distance[\indexLED\indexTransmitter] = \norm[{\incidenceVector[\indexLED\indexTransmitter]}]_2~,
\end{align}
\begin{align}
\cos(\angleLOSandTX[\indexLED\indexTransmitter]) = \frac{\directionLED[{\indexLED\indexTransmitter}]^{\rm T}(\locationDetector- \locationLED[{\indexLED\indexTransmitter}])}{\distance[\indexLED\indexTransmitter]}
=
\frac{\incidenceVector[\indexLED\indexTransmitter]^{\rm T}\directionLED[\indexLED\indexTransmitter]}{\norm[{\incidenceVector[\indexLED\indexTransmitter]}]_2}~,
\end{align}
and
\begin{align}
\cos(\angleLOSandRX[\indexLED\indexTransmitter]) = \frac{\directionDetector^{\rm T}(\locationLED[{\indexLED\indexTransmitter}]-\locationDetector )}{\distance[\indexLED\indexTransmitter]}
=-
\frac{\incidenceVector[\indexLED\indexTransmitter]^{\rm T}\directionDetector}{\norm[{\incidenceVector[\indexLED\indexTransmitter]}]_2}~.
\end{align}
Therefore \eqref{eq:power} can be rewritten as
\begin{align}
\hspace{-3mm}
\power[\indexLED\indexTransmitter]=-\frac{(\mode+1)\Ar}{2\pi}
\rect[{\frac{\angleLOSandRX[\indexLED\indexTransmitter]}{\FOV}}] \rect[{\frac{\angleLOSandTX[\indexLED\indexTransmitter]}{\pi/2}}] \times\function[{\incidenceVector[\indexLED\indexTransmitter]}]
~,
\label{eq:powerVector}
\end{align}
where
\begin{align}
\function[{\incidenceVector[\indexLED\indexTransmitter]}]
=&
\frac{(\incidenceVector[\indexLED\indexTransmitter]^{\rm T}\directionLED[\indexLED\indexTransmitter])^{\mode}\incidenceVector[\indexLED\indexTransmitter]^{\rm T}\directionDetector}{\norm[{\incidenceVector[\indexLED\indexTransmitter]}]_2^{\mode+3}}
\nonumber \\=&
(
\incidenceElements[\indexLED\indexTransmitter][a]\directionLEDElements[\indexLED\indexTransmitter][x]
+
\incidenceElements[\indexLED\indexTransmitter][b]\directionLEDElements[\indexLED\indexTransmitter][y]
+
\incidenceElements[\indexLED\indexTransmitter][c]\directionLEDElements[\indexLED\indexTransmitter][z]
)^{\mode}\nonumber \\&~~~~~~~~~\times
\frac{
(
\incidenceElements[\indexLED\indexTransmitter][a]\directionDetectorElements[x]
+
\incidenceElements[\indexLED\indexTransmitter][b]\directionDetectorElements[y]
+
\incidenceElements[\indexLED\indexTransmitter][c]\directionDetectorElements[z]
)
}{(\incidenceElements[\indexLED\indexTransmitter][a]^2+\incidenceElements[\indexLED\indexTransmitter][b]^2+\incidenceElements[\indexLED\indexTransmitter][c]^2)^{\frac{\mode+3}{2}}}
\label{eq:function}~.
\end{align}

It is worth noting that \eqref{eq:powerVector} includes only the \ac{LOS} component, i.e., first order Lambertian pattern, since the \ac{LOS} component of signal is dominant compared to other multipath components for \ac{VLC} systems \cite{Barry_1993,zhou_2012}. Hence, we discuss the theoretical analyses given in the following sections based on \ac{LOS} component. In this study, we assume that all \acp{LED} in \acp{VAP} are identical. In addition to that, the locations of the transmitters and directions of each \ac{LED} are assumed to be known at the receiver, which can be shared through periodic broadcast messages. The unique identity of each \ac{LED} transmitter and its corresponding \ac{VAP} are assumed to be decodable at the receiver, which can be achieved by assigning different codes/headers from each \ac{LED} transmitter. Accordingly, it is assumed that the receiver is able to measure the \ac{RSS} associated with each \ac{LED} transmitter.

\section{AOA-Based Localization}
\label{sec:AOA}
In \ac{AOA}-based localization, a \ac{VLC} receiver selects one of the \ac{LED} transmitters for each \ac{VAP} based on the RSS information and locates itself by exploiting the orientations of \ac{LED} transmitters. Geometrically, \ac{AOA}-based localization relies on finding a point in 3D space such that it minimizes the sum of distances (or squared distances) between this point and all the other lines extended from the directions of \ac{LED} transmitters selected by the receiver \cite{Bilgi_2012,eroglu_2015}. This approach can allow the receiver to locate itself with only two anchor nodes when the receiver is close to the intersection of the corresponding \ac{LED} directions. 

\begin{figure}[!t]
\centering
\includegraphics[width =2.6in]{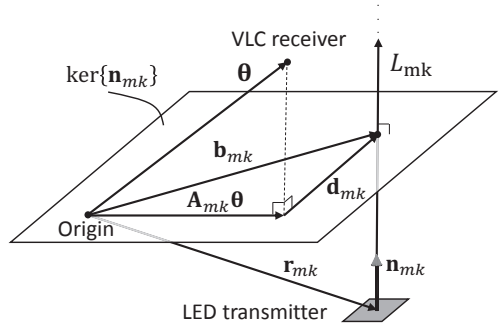}
\label{fig:AOAdistance}
\caption{The distance between VLC receiver's location and the line extended by the direction of an LED transmitter.}
\label{fig:AOA}
\end{figure}

In order to analyze the \ac{AOA}-based localization, let $\lineLED[\indexLED\indexTransmitter]$ be the line associated with the $\indexLED$th \ac{LED} transmitter of $\indexTransmitter$th \ac{VAP} as illustrated in \figurename~\ref{fig:AOA},  where its origin and direction are captured by the vectors  $\locationLED[{\indexLED\indexTransmitter}]$ and $\directionLED[{\indexLED\indexTransmitter}]$, respectively. In addition, let $\projectionlineLEDNull[{\indexLED\indexTransmitter}]\in\realNumbers^{3\times3}$ be the projection matrix which projects every vector $\anObservationVector\in\realNumbers^{3\times1}$ onto the null space of $\directionLED[{\indexLED\indexTransmitter}]$, i.e., $\kernel[{\directionLED[{\indexLED\indexTransmitter}]}]$, which can be calculated by
$
\projectionlineLEDNull[{\indexLED\indexTransmitter}] = \identityMatrix[3]-\directionLED[{\indexLED\indexTransmitter}]\directionLED[{\indexLED\indexTransmitter}]^{\rm T}
$. 
Hence, any vector in the column space of $\projectionlineLEDNull[{\indexLED\indexTransmitter}]$ is orthogonal to the direction of $\lineLED[\indexLED\indexTransmitter]$. In particular, the projection of any point on $\lineLED[\indexLED\indexTransmitter]$ yields the same point, i.e., the intersection between $\lineLED[\indexLED\indexTransmitter]$ and the subspace spanned by the columns of $\projectionlineLEDNull[{\indexLED\indexTransmitter}]$. The intersection point $\intersectionPoint[\indexLED\indexTransmitter]$ can be simply calculated as $\projectionlineLEDNull[{\indexLED\indexTransmitter}]\locationLED[{\indexLED\indexTransmitter}]$ since $\locationLED[{\indexLED\indexTransmitter}]$ is a point on $\lineLED[\indexLED\indexTransmitter]$. Hence, the distance vector between an arbitrary point $\parameterVector\in\realNumbers^{3\times 1}$ and $\lineLED[\indexLED\indexTransmitter]$
is given by $\distanceVector[\indexLED\indexTransmitter] = \intersectionPoint[\indexLED\indexTransmitter] - \projectionlineLEDNull[{\indexLED\indexTransmitter}] \parameterVector$, which yields
\begin{align}
\intersectionPoint[\indexLED\indexTransmitter] = \projectionlineLEDNull[{\indexLED\indexTransmitter}] \parameterVector+\distanceVector[\indexLED\indexTransmitter]~.
\end{align}
By stacking the set of equations related to \ac{LED} transmitters, we obtain
\begin{align}
\intersectionPoint[] = \projectionlineLEDNull[] \parameterVector+\distanceVector[]~,
\end{align}
where 
\begin{align}
\intersectionPoint[] = 
\begin{bmatrix}
\intersectionPoint[{\setSelectLED[1]}1]\\
\vdots\\
\intersectionPoint[{\setSelectLED[\indexTransmitter]}\indexTransmitter]\\
\vdots\\
\intersectionPoint[{\setSelectLED[\numberOfTransmitters]}\numberOfTransmitters]\\
\end{bmatrix},~
\projectionlineLEDNull[] = 
\begin{bmatrix}
\projectionlineLEDNull[{\setSelectLED[1]}1]\\
\vdots\\
\projectionlineLEDNull[{\setSelectLED[\indexTransmitter]}\indexTransmitter]\\
\vdots\\
\projectionlineLEDNull[{\setSelectLED[\numberOfTransmitters]}\numberOfTransmitters]\\
\end{bmatrix},~
\distanceVector[] = 
\begin{bmatrix}
\distanceVector[{\setSelectLED[1]}1]\\
\vdots\\
\distanceVector[{\setSelectLED[\indexTransmitter]}\indexTransmitter]\\
\vdots\\
\distanceVector[{\setSelectLED[\numberOfTransmitters]}\numberOfTransmitters]\\
\end{bmatrix}~,
\end{align}
and $\setSelectLED[\indexTransmitter]$ is the index of selected \ac{LED} transmitter of $\indexTransmitter$th \ac{VAP}. Therefore the objective function which minimizes the sum  of squared error is solved via an \ac{LS} estimator given by
 \begin{align}
\locationDetectorEstimate &= {\projectionlineLEDNull[{}]^{\rm \dagger}}\intersectionPoint[]~.
\label{eq:CBsolutionLS}
\end{align}

\def\weights[#1]{\beta_{#1}}
\def\varianceOperator[#1]{{{\rm var}\left(#1\right)}}
\subsection{Weighting LED Transmitters}
Treating the \ac{AOA} information obtained from each of the  \ac{LED} transmitters identically, as in \eqref{eq:CBsolutionLS}, may cause significant positioning error since the distance between exact position of the receiver and the line extended by the direction of \ac{LED} transmitter may be large in some cases. In other words, less reliable \ac{AOA} information may bias the location estimation and degrade localization accuracy. In order to address this issue, we consider an objective function which minimizes the weighted sum of squared distance as
\begin{align}
\locationDetectorEstimate &=  
\arg\min_{\parameterVector}\sum_{\substack{\indexTransmitter=1\\\indexLED=\setSelectLED[\indexTransmitter]}}^\numberOfTransmitters
\weights[\indexLED\indexTransmitter]\norm[{\distanceVector[\indexLED\indexTransmitter] }]_2^2\nonumber 
\\&=
\arg\min_{\parameterVector}\sum_{\substack{\indexTransmitter=1\\\indexLED=\setSelectLED[\indexTransmitter]}}^\numberOfTransmitters
\weights[\indexLED\indexTransmitter]\norm[{\projectionlineLEDNull[{\indexLED\indexTransmitter}] \parameterVector-\intersectionPoint[\indexLED\indexTransmitter]}]_2^2~,
\label{eq:connectivity2}
\end{align}
where $\weights[\indexLED\indexTransmitter]$ is the weighing factor for the distance between  $\parameterVector$ and $\lineLED[\indexLED\indexTransmitter]$. The purpose of weighing factors is to  take the variability of $\distanceVector[\indexLED\indexTransmitter] $ into account in the optimization. Under the assumptions of $\expectedOperator[{\distanceVector[\indexLED\indexTransmitter]}]=0$ and $\expectedOperator[{\distanceVector[nl]^{\rm T}}{\distanceVector[\indexLED\indexTransmitter]}]=0$, according to Gauss-Markov theorem \cite{Kay_1993}, the optimal weighting factors that yield minimum variance estimator for \eqref{eq:connectivity2} is obtained as 
\begin{align}
\weights[\indexLED\indexTransmitter] =c_0 \times \frac{1}{\expectedOperator[{\norm[{\distanceVector[\indexLED\indexTransmitter]}]^2}]}~.
\label{eq:varianceDistance}
\end{align}
where $c_0$ is an arbitrary positive real number. In order to calculate \eqref{eq:varianceDistance}, the statistical characteristics of ${\distanceVector[\indexLED\indexTransmitter]}$ should be known {\em a priori}. To this end,  we exploit \ac{RSS} information associated with \ac{LED} transmitters in this study.

\begin{figure}[!t]
\centering
\includegraphics[width=3.3in]{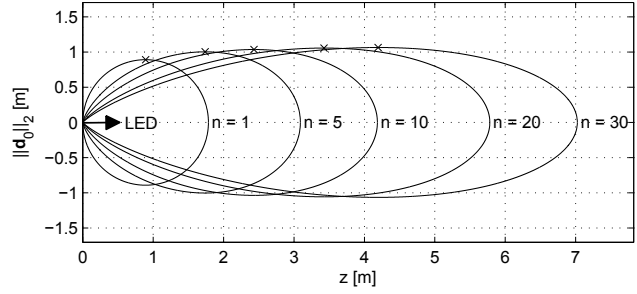}
\caption{RSS contours for a given LED mode. All the locations on the RSS contours have identical RSS values.}
\label{fig:AOAdist}
\end{figure}

Without loss of generality, consider an \ac{LED} transmitter located at the origin and its orientation is set to $\directionLED[0] = [0,0,1]^{\rm T}$. Assuming that the \ac{LED} transmitter and the \ac{VLC} receiver face each other, i.e., $\directionDetector = -\directionLED[0]$, $\FOV=\pi/2$, and ${\angleLOSandTX[\indexLED\indexTransmitter]}\le{\pi/2}$, we rewrite \eqref{eq:powerVector} as
\begin{align}
\power[0]=\frac{(\mode+1)\Ar}{2\pi}\frac{
z^{n+1}
}{(x^2+y^2+z^2)^{\frac{\mode+3}{2}}}
~.
\label{eq:powerVectorCase}
\end{align}
We  then obtain the identity which characterizes the distance between the \ac{LED} direction and \ac{VLC} receiver's location  as
\def\constant{C}
\begin{align}
\norm[{\distanceVector[0] }]_2^2
=\constant^{{{\frac{2}{\mode+3}}}} z^{\frac{2\mode+2}{\mode+3}}-{z^2}
~,
\label{eq:isohypse}
\end{align}
where $\norm[{\distanceVector[0] }]_2^2=x^2+y^2$ and $\constant \triangleq  \frac{(\mode+1)\Ar z}{2\pi\power[0]} $. It is worth noting that \eqref{eq:isohypse} also identifies the locations where \ac{VLC} receiver could observe the same \ac{RSS} and forms an \ac{RSS} contour. For example, when $\power[0]$ is fixed to $0.1$~mW and $\Ar=1$~cm$^2$, the potential locations of a \ac{VLC} receiver are shown in \figurename~\ref{fig:AOAdist} for a given \ac{LED} mode. Under the assumption that the receiver is equally likely to be on the locations that satisfy \eqref{eq:isohypse}, one should obtain $\expectedOperator[{\norm[{\distanceVector[0] }]_2^2}]$ in order to find optimal weights given in \eqref{eq:varianceDistance}. However, the calculation of $\expectedOperator[{\norm[{\distanceVector[0] }]_2^2}]$ is not trivial due to nonlinear characteristics of \eqref{eq:isohypse}. On the other hand, considering the narrow beamwidth of the \ac{LED} with high modes,  \ac{RSS} contours are spread around on $z$-axis as shown in \figurename~\ref{fig:AOAdist}. Therefore, it becomes more likely that the \ac{VLC} receiver location far away from the \ac{LED} direction. Hence, in this study, we approximate the exact distribution of $\norm[{\distanceVector[0] }]_2$ with a uniform \ac{PDF} given by
\def\dMax{d_{\rm max}}
\begin{align}
\probability[{\norm[{\distanceVector[0] }]_2}] =  
\begin{cases}
	\frac{1}{2\pi\dMax}, & \norm[{\distanceVector[0] }]_2=\dMax \\
	0, & {\rm otherwise} \\
\end{cases}
\end{align}
where 
\def\constantB{B}
\begin{align}
\dMax=
\max\left({\norm[{\distanceVector[0] }]_2}\right)= \sqrt{\constantB^{-\frac{n+1}{2}}\constant- \constantB^{-\frac{n+3}{2}}\constant}
\end{align}
and $\constantB \triangleq (2\mode+6)/(2\mode+2)$. Therefore the variance of $\norm[{\distanceVector[0] }]_2$ is obtained as $ \dMax^2$ approximately. By induction, the optimum weights in \eqref{eq:varianceDistance} are derived as
\begin{align}
\weights[\indexLED\indexTransmitter]  \approx c_0\times\frac{\frac{2\pi\power[\indexLED\indexTransmitter]}{(\mode+1)\Ar z}}{{\constantB^{-\frac{n+1}{2}}- \constantB^{-\frac{n+3}{2}}}}\stackrel{(a)}{=}\power[\indexLED\indexTransmitter]~,
\label{eq:weights}
\end{align}
where (a) follows from the appropriate selection of $c_0$. As a result of \eqref{eq:weights}, \ac{LED} transmitters are weighted based on their \ac{RSS} information $\power[\indexLED\indexTransmitter]$.

\setcounter{equation}{28}
\begin{figure*}[b]
\begin{align}
\begin{bmatrix}
\frac{\partial\power[\indexLED\indexTransmitter]}{\partial \incidenceElements[\indexLED\indexTransmitter][a]}\\
\frac{\partial\power[\indexLED\indexTransmitter]}{\partial \incidenceElements[\indexLED\indexTransmitter][b]}\\
\frac{\partial\power[\indexLED\indexTransmitter]}{\partial \incidenceElements[\indexLED\indexTransmitter][c]}\\
\end{bmatrix}^{\rm T}
\textrm{=}
\begin{bmatrix}
\textrm{-}\frac{(\mode+1)\Ar}{2\pi}
\rect[{\frac{\angleLOSandRX[\indexLED\indexTransmitter]}{\FOV}}]\rect[{\frac{\angleLOSandTX[\indexLED\indexTransmitter]}{\pi/2}}] 
\derivativeFunction[{\incidenceElements[\indexLED\indexTransmitter][a]}][{\directionLEDElements[\indexLED\indexTransmitter][x]}][{\directionDetectorElements[x]}][{\incidenceElements[\indexLED\indexTransmitter][b]\directionLEDElements[\indexLED\indexTransmitter][y]\textrm{+}\incidenceElements[\indexLED\indexTransmitter][c]\directionLEDElements[\indexLED\indexTransmitter][z]}][{\incidenceElements[\indexLED\indexTransmitter][b]\directionDetectorElements[y]\textrm{+}\incidenceElements[\indexLED\indexTransmitter][c]\directionDetectorElements[z]}][{\incidenceElements[\indexLED\indexTransmitter][b]^2\textrm{+}\incidenceElements[\indexLED\indexTransmitter][c]^2}][\mode]\\
\textrm{-}\frac{(\mode+1)\Ar}{2\pi}
\rect[{\frac{\angleLOSandRX[\indexLED\indexTransmitter]}{\FOV}}]\rect[{\frac{\angleLOSandTX[\indexLED\indexTransmitter]}{\pi/2}}] 
\derivativeFunction[{\incidenceElements[\indexLED\indexTransmitter][b]}][{\directionLEDElements[\indexLED\indexTransmitter][y]}][{\directionDetectorElements[y]}][{\incidenceElements[\indexLED\indexTransmitter][a]\directionLEDElements[\indexLED\indexTransmitter][x]\textrm{+}\incidenceElements[\indexLED\indexTransmitter][c]\directionLEDElements[\indexLED\indexTransmitter][z]}][{\incidenceElements[\indexLED\indexTransmitter][a]\directionDetectorElements[x]\textrm{+}\incidenceElements[\indexLED\indexTransmitter][c]\directionDetectorElements[z]}][{\incidenceElements[\indexLED\indexTransmitter][a]^2\textrm{+}\incidenceElements[\indexLED\indexTransmitter][c]^2}][\mode]\\
\textrm{-}\frac{(\mode+1)\Ar}{2\pi}
\rect[{\frac{\angleLOSandRX[\indexLED\indexTransmitter]}{\FOV}}]\rect[{\frac{\angleLOSandTX[\indexLED\indexTransmitter]}{\pi/2}}] 
\derivativeFunction[{\incidenceElements[\indexLED\indexTransmitter][c]}][{\directionLEDElements[\indexLED\indexTransmitter][z]}][{\directionDetectorElements[z]}][{\incidenceElements[\indexLED\indexTransmitter][a]\directionLEDElements[\indexLED\indexTransmitter][x]\textrm{+}\incidenceElements[\indexLED\indexTransmitter][b]\directionLEDElements[\indexLED\indexTransmitter][y]}][{\incidenceElements[\indexLED\indexTransmitter][a]\directionDetectorElements[x]\textrm{+}\incidenceElements[\indexLED\indexTransmitter][b]\directionDetectorElements[y]}][{\incidenceElements[\indexLED\indexTransmitter][a]^2\textrm{+}\incidenceElements[\indexLED\indexTransmitter][b]^2}][\mode]\\
\end{bmatrix}^{\rm T}
\label{eq:closedForm}
\end{align}
\vspace{-5mm}
\end{figure*}
\setcounter{equation}{18}

The problem given in \eqref{eq:connectivity2} corresponds to an unconstrained quadratic optimization problem, which can be solved via \ac{LS} method. The closed-form solution of \eqref{eq:connectivity2} can then be obtained as
\begin{align}
\locationDetectorEstimate &= {\projectionlineLEDNull[{\rm w}]}^{\rm \dagger}\intersectionPoint[{\rm w}]~,
\label{eq:CBsolution}
\end{align}
where $\displaystyle
\projectionlineLEDNull[{{\rm w}}]=\sum_{\substack{\indexTransmitter=1\\\indexLED=\setSelectLED[\indexTransmitter]}}^\numberOfTransmitters \power[\indexLED\indexTransmitter]\projectionlineLEDNull[{\indexLED\indexTransmitter}]
$
and
$\displaystyle
\intersectionPoint[{\rm w}]=\sum_{\substack{\indexTransmitter=1\\\indexLED=\setSelectLED[\indexTransmitter]}}^\numberOfTransmitters\power[\indexLED\indexTransmitter]\intersectionPoint[\indexLED\indexTransmitter]
$.

\section{RSS-Based Localization}
\label{sec:RSSI}
Under the assumption of availability of physical characteristics of \acp{LED}, the receiver can locate its own location based on \ac{RSS} information associated with  \ac{LED} transmitters. Let $\observationVector\in\realNumbers^{\numberOfLEDs\numberOfTransmitters\times1}$ be the observation vector which is given by
\begin{align}
\observationVector = \powerVector[{\parameterVector}] + \noiseVector~,
\label{eq:observation}
\end{align}
where  $\noiseVector\in\realNumbers^{\numberOfTransmitters\numberOfLEDs\times1}\sim \realGaussian[0][{\noiseVariance\identityMatrix[\numberOfTransmitters\numberOfLEDs]}]$ is an additive Gaussian noise vector, $\powerVector[{\parameterVector}]=\vectorization[\powerMatrix]\in\realNumbers^{\numberOfLEDs\numberOfTransmitters\times1}$, and $\powerMatrix\in\realNumbers^{\numberOfLEDs\times\numberOfTransmitters}$ is a matrix which contains the exact \ac{RSS} information with the entry on $\indexLED$th row and $\indexTransmitter$th column being $\power[\indexLED\indexTransmitter]$. The log-likelihood function for the location of \ac{VLC} receiver is then expressed as
\begin{align}
\loglikelihood[\parameterVector] = \log \probability[\observationVector;\parameterVector]~,
\label{eq:likelihood}
\end{align}
where
\begin{align}
\probability[\observationVector;\parameterVector]= \frac{1}{2\pi\noiseVariance}\exp{\left(-\frac{1}{2\noiseVariance}  (\observationVector - \powerVector[{\parameterVector}])^{\rm T}(\observationVector - \powerVector[{\parameterVector}]) \right)}
\end{align}
and $\parameterVector\in\realNumbers^{3\times1}$ is the parameter vector which corresponds to the location of the \ac{VLC} receiver, i.e., $\locationDetector$. \ac{ML} estimate of $\locationDetector$ is therefore formulated as
\begin{align}
\locationDetectorEstimate &=  \arg\max_{\parameterVector}\loglikelihood[\parameterVector]\propto \arg\max_{\parameterVector}{\left( - (\observationVector - \powerVector[{\parameterVector}])^{\rm T}(\observationVector - \powerVector[{\parameterVector}]) \right)}~.
\label{eq:ML}
\end{align}
As a result, \eqref{eq:ML} can be expressed as a \ac{NLS} problem given by
\begin{align}
\locationDetectorEstimate = \arg\min_{\parameterVector}
{\left(\norm[\observationVector - {\powerVector[{\parameterVector}]}]_2^2 \right)}~.
\label{eq:NLS}
\end{align}

In the following subsections, first, we introduce a learning rule in order to solve \eqref{eq:NLS}. We then investigate the non-convex structure of \eqref{eq:NLS} and address the initialization issues of the learning rule via the \ac{RRC} algorithm. Finally, we provide \ac{CRLB} of \eqref{eq:NLS}.

\subsection{Learning Rule for RSS-Based Localization}
In order to solve the system of nonlinear equations in \eqref{eq:NLS}, we follow multivariate Newton-Raphson method, which yields
\def\stepSize{\eta}
\begin{align}
\parameterVector^{i+1} = \parameterVector^{i} - \stepSize\jacobianMatrix^{\dagger}(\observationVector - \powerVector[{\parameterVector^{i}}])~,
\label{eq:update}
\end{align}
where $\stepSize\in(0,1]$ is the step size and $\jacobianMatrix$ is the Jacobian matrix of $\powerVector[{\parameterVector}]$ with respect to $\parameterVector$. As $\theta_1$, $\theta_2$, and $\theta_3$ corresponds to $x$, $y$, and $z$ in \eqref{eq:observation}, respectively, $\jacobianMatrix$ can be explicitly given by
\begin{align}
\jacobianMatrix = 
\begin{bmatrix}
\frac{\partial\power[00]}{\partial x} &
\frac{\partial\power[00]}{\partial y} &
\frac{\partial\power[00]}{\partial z}\\
\frac{\partial\power[10]}{\partial x} &
\frac{\partial\power[10]}{\partial y} &
\frac{\partial\power[10]}{\partial z}\\
\vdots & 
\vdots &
\vdots \\
\frac{\partial\power[\numberOfLEDs\numberOfTransmitters]}{\partial x} &
\frac{\partial\power[\numberOfLEDs\numberOfTransmitters]}{\partial y} &
\frac{\partial\power[\numberOfLEDs\numberOfTransmitters]}{\partial z}\\
\end{bmatrix}~.
\label{eq:jacobian}
\end{align}
As it can be seen in \eqref{eq:jacobian}, each row of $\jacobianMatrix$ indicates how the \ac{RSS} associated with an \ac{LED} transmitter change when the receiver moves in one of the axes, i.e., $x$, $y$, and $z$. Considering the chain rule of derivatives, the row associated with the $\indexLED$th \ac{LED} transmitter of $\indexTransmitter$th \ac{VAP} is calculated as
\begin{align}
\begin{bmatrix}
\frac{\partial\power[\indexLED\indexTransmitter]}{\partial x}\\
\frac{\partial\power[\indexLED\indexTransmitter]}{\partial y}\\
\frac{\partial\power[\indexLED\indexTransmitter]}{\partial z}\\
\end{bmatrix}^{\rm T}
=
\begin{bmatrix}
\frac{\partial\power[\indexLED\indexTransmitter]}{\partial \incidenceElements[\indexLED\indexTransmitter][a]}\\
\frac{\partial\power[\indexLED\indexTransmitter]}{\partial \incidenceElements[\indexLED\indexTransmitter][b]}\\
\frac{\partial\power[\indexLED\indexTransmitter]}{\partial \incidenceElements[\indexLED\indexTransmitter][c]}\\
\end{bmatrix}^{\rm T}
\begin{bmatrix}
\frac{\partial\incidenceElements[\indexLED\indexTransmitter][a]}{\partial x} & \frac{\partial\incidenceElements[\indexLED\indexTransmitter][a]}{\partial y} &
\frac{\partial\incidenceElements[\indexLED\indexTransmitter][a]}{\partial z}\\
\frac{\partial\incidenceElements[\indexLED\indexTransmitter][b]}{\partial x} & \frac{\partial\incidenceElements[\indexLED\indexTransmitter][b]}{\partial y} &
\frac{\partial\incidenceElements[\indexLED\indexTransmitter][b]}{\partial z}\\
\frac{\partial\incidenceElements[\indexLED\indexTransmitter][c]}{\partial x} & \frac{\partial\incidenceElements[\indexLED\indexTransmitter][c]}{\partial y} &
\frac{\partial\incidenceElements[\indexLED\indexTransmitter][c]}{\partial z}\\
\end{bmatrix}~.
\label{eq:chain}
\end{align}
Since $\incidenceVector[\indexLED\indexTransmitter]=\locationDetector- \locationLED[{\indexLED\indexTransmitter}]$, the matrix in \eqref{eq:chain}, i.e., Jacobian of $\incidenceVector[\indexLED\indexTransmitter]$ respect to $\parameterVector$, becomes an identity matrix. Therefore, \eqref{eq:chain} can be directly calculated by evaluating the derivative of $\power[\indexLED\indexTransmitter]$  with respect to $\incidenceVector[\indexLED\indexTransmitter]$  at $\incidenceVector[\indexLED\indexTransmitter]=\locationDetector- \locationLED[{\indexLED\indexTransmitter}]$. 
\color{black}The derivative of $\power[\indexLED\indexTransmitter]$  with respect to $\incidenceVector[\indexLED\indexTransmitter]$ is analytically given  in \eqref{eq:closedForm} by employing an auxiliary function $\derivativeFunction[x][a][b][k][l][m][n]$ defined as
\begin{align}
\derivativeFunction[x][a][b][k][l][m][n] &\triangleq \frac{\partial \function[x;a,b,k,l,m,n]}{\partial x} \nonumber \\
=&\frac{(ax+k)^{(n-1)}(c_3x^3+c_2x^2+c_1x+c_0)}{(x^2+m)^{\frac{n+5}{2}}}~,
\nonumber
\end{align}
where $\function[x;a,b,k,l,m,n]$ is a function based on the structure of $\power[\indexLED\indexTransmitter]$ and keeping one of the elements of incidence vector $\incidenceVector[\indexLED\indexTransmitter]$ as a variable and the others as constants in \eqref{eq:function}. It is explicitly defined as
\begin{align}
\function[x;a,b,k,l,m,n] \triangleq \frac{(ax+k)^{(n-1)}(bx+l)}{(x^2+m)^{\frac{n+3}{2}}},
\end{align}
where $c_0 = kbm+lamn$, $c_1 = bmna+bma-lk(n+3)$, $c_2 = -(kb(n+2)+3la)$, and $c_3 = -2ab$.
\color{black}
 After plugging $\incidenceVector[\indexLED\indexTransmitter]=\locationDetector- \locationLED[{\indexLED\indexTransmitter}]$ into \eqref{eq:closedForm}, i.e., $\incidenceElements[\indexLED\indexTransmitter][a]=x-x_{\indexLED\indexTransmitter}$, $\incidenceElements[\indexLED\indexTransmitter][b]=y-y_{\indexLED\indexTransmitter}$, and $\incidenceElements[\indexLED\indexTransmitter][c]=z-z_{\indexLED\indexTransmitter}$,  \eqref{eq:jacobian} and \eqref{eq:chain} are obtained theoretically. Therefore, the learning rule given in \eqref{eq:update} can be calculated analytically, which eliminates the numerical calculation of $\jacobianMatrix$ and yields less complex structures.

\subsection{Non-Convex Structure of RSS-Based Localization}
\begin{figure}[!t]
\centering
{\includegraphics[width =3.5in]{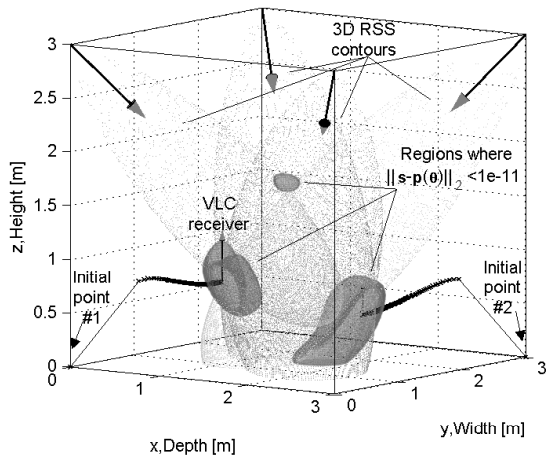}
}
\caption{Convexity of \eqref{eq:NLS}. In this setup, there are three regions where the modeling error is less than $1$e-$11$ and lead to three different local optimum points which indicate non-convex structure of  \eqref{eq:NLS}. While the first initial point allows \eqref{eq:update} to find the global optimum, i.e., receiver's location, the second initial point  converges to one of the local optima. }
\label{fig:nonConvex}
\end{figure} 

Although Lambertian pattern offers a convex set by itself, \eqref{eq:NLS} is not a convex function in general. This is due to the fact that the set of feasible solutions associated with each \ac{LED} transmitter may become closer to each other in multiple locations in 3D geometry. For instance, consider a $3$~m $\times$ $3$~m $\times$ $3$~m where the \acp{VAP} are located at the corners as illustrated in \figurename~\ref{fig:nonConvex}. In this setup, we assume that each \ac{VAP} has single \ac{LED} transmitter where $\mode = 30$ and the angles between LED directions and side walls are set to $45$ degrees. The \ac{VLC} receiver parameters are set as $\FOV=85$ degrees, $\Ar=1$~cm$^2$, $\locationDetector=[1,1,0.75]^{\rm T}$, and $\directionDetector=[0,0,1]^{\rm T}$. Having individual \ac{RSS} information from each \ac{LED} transmitter based on \eqref{eq:power}, we then find the locations where \ac{VLC} receiver could observe the same \ac{RSS} information for each \ac{LED} transmitter in 3D geometry, which corresponds to \ac{RSS} contours. 

As it can been seen in \figurename~\ref{fig:nonConvex}, the \ac{VLC} receiver is located at the intersection of \ac{RSS} contours. However, there are other locations in the room where the \ac{RSS} contours get closer to each other. When we evaluate \eqref{eq:NLS} and find the locations where the modeling error, i.e., $\norm[\observationVector - {\powerVector[{\parameterVector}]}]_2$, is less than $1$e-$11$, we observe three different regions which indicate at least three different local optimum points. Therefore, different initial points for the learning rule given in 
\eqref{eq:update} may converge to different location estimates. For example, in \figurename~\ref{fig:nonConvex}, while the first initial point  yields the true location of the  \ac{VLC} receiver, the second one converges to one of the local optima, i.e., the point where the \ac{RSS} contours get closer to each other.

\subsection{Initialization of RSS-Based Localization}
In order to increase the likelihood that \eqref{eq:update} converges to the global optimum, one may use the solution of \ac{AOA}-based localization method  as an initial point. The \ac{AOA}-based solution is useful, such that it provides a closed-form solution that is close to the exact location of the \ac{VLC} receiver, and hence does not require exhaustive search methods. However, this method may still lead to wrong results in some cases.  In order to identify better initial points which increase the likelihood of finding the global optimum, i.e., location of the receiver, we propose a heuristic algorithm which is based on random search \cite{Tao_2008} and unsupervised clustering methods \cite{hastie01statisticallearning}, which we call as the \ac{RRC} algorithm.

\def\numberOfRandomPoints{S}
\def\numberOfselectPoints{G}
\def\aPromisingPoint[#1]{{\rm \bf p}_{#1}}

\ac{RRC} algorithm includes two steps: 1) random reporting and 2) clustering. In the first step, the algorithm generates $\numberOfRandomPoints$ random location samples in 3D space. The rewards achieved by the location samples are  then calculated based on the modeling error in \eqref{eq:likelihood}.  By evaluating the rewards, only  $\numberOfselectPoints<\numberOfRandomPoints$ points that give the most promising results are chosen and the rest of the sample locations are eliminated. If the solution space is sampled in a sufficiently dense way, the favorable location samples should be located in the space as {\em clusters} since the objective function is non-convex and local regions appear in 3D space as clusters as illustrated in \figurename~\ref{fig:nonConvex}. 

\def\iteration{j}
\def\numberOfClusters{C}
\def\centerOfCluster[#1]{{\rm \bf c}_{#1}}
\def\setR[#1][#2]{R_{\it #1}^{#2}}

\def\cardinalityOperator[#1]{
\left\vert{#1}\right\vert
}

\setcounter{equation}{28}

In the second step, we aim at finding the centroids of the clusters. The reason for finding the centroids is that they are more likely to be around the true location of the \ac{VLC} receiver, which make them good candidates for the initial points. To this end, we exploit unsupervised learning methods which are well investigated in the machine learning literature. Due to its simplicity, in this study, we employ $K$-means algorithm which is an effective method for finding the centroids of the {clusters} in a high-dimensional space \cite{hastie01statisticallearning,aharon_2006,gersho1992vector}. Essentially, $K$-means algorithm converges to the centroids of the clusters with iterations.  
It applies two steps per each iteration. In the first step of $\iteration$th iteration of  $K$-means algorithm, the points are partitioned into $\numberOfClusters$ clusters, {$\{ \setR[i=1][\iteration], \setR[i=2][\iteration], \dots, \setR[i=\numberOfClusters][\iteration]  \}$}, based on their distances to the centroids obtained in $(\iteration-1)$th iteration. In the second step, the $i$th centroid, i.e., $\centerOfCluster[i]$, is updated to better fit for the location samples as
\begin{align}
\centerOfCluster[i]^{(j+1)} = \frac{1}{
\cardinalityOperator[{\setR[i][(\iteration)]}]
} \sum_{\ell\in \setR[i][(\iteration)]} \aPromisingPoint[\ell]~,
\end{align}
where $\aPromisingPoint[\ell]$ is the $\ell$th point obtained from random reporting step and $\cardinalityOperator[\cdot]$ is the cardinality of its argument. 
Once we obtain the centroids, we utilize them as initial points for the update rule in \eqref{eq:update}. We then select the best point based on \eqref{eq:NLS}.

\subsection{Cram\'{e}r-Rao Lower Bound  for RSS-Based Localization}
Without loss of generality, when $\anObservationVector\sim \realGaussian[\meanFunctionOfParameters][\coVarianceFunctionOfParameters]$, the element on $i$th row and $j$th column of \ac{FIM} $\FisherInformationMatrix$ can be calculated as \cite{Kay_1993}
\begin{align}
[\FisherInformationMatrix]_{ij} = &\left[\frac{\partial\meanFunctionOfParameters}{\partial \theta_i}\right]^{\rm T}\coVarianceFunctionOfParameters^{-1} \left[\frac{\partial\meanFunctionOfParameters}{\partial \theta_j}\right]\nonumber \\&
+\frac{1}{2}{\rm tr}\left[\coVarianceFunctionOfParameters^{-1} 
\frac{\partial\coVarianceFunctionOfParameters}{\partial \theta_i}\coVarianceFunctionOfParameters^{-1} 
\frac{\partial\coVarianceFunctionOfParameters}{\partial \theta_j}
 \right]~.
\end{align}
Since the noise term in \eqref{eq:observation} is assumed to be white Gaussian noise and does not depend on the location of receiver, the second term is zero and first term yields
$
[\FisherInformationMatrix]_{ij} = \frac{1}{\noiseVariance}\left[\frac{\partial\meanFunctionOfParameters}{\partial \theta_i}\right]^{\rm T}\left[\frac{\partial\meanFunctionOfParameters}{\partial \theta_j}\right]
$.
Based on \eqref{eq:observation}, $\meanFunctionOfParameters$ corresponds to mean power captured power from all transmitters, i.e., $\meanFunctionOfParameters=\powerVector[{\parameterVector}]$, which leads to the following expression:
\begin{align}
[\FisherInformationMatrix]_{ij} = &\frac{1}{\noiseVariance}\left[\frac{\partial\powerVector[{\parameterVector}]}{\partial \theta_i}\right]^{\rm T}\left[\frac{\partial\powerVector[{\parameterVector}]}{\partial \theta_j}\right]~.
\label{eq:FIMelements}
\end{align}
By using \eqref{eq:FIMelements}, complete \ac{FIM} can be then calculated as
\begin{align}
\FisherInformationMatrix = &\frac{1}{\noiseVariance}\jacobianMatrix^{\rm T}\jacobianMatrix~,
\end{align}
where $\jacobianMatrix$ is the Jacobian matrix of $\powerVector[{\parameterVector}]$ with respect to $\parameterVector$, which is already  derived {\em analytically} by using \eqref{eq:chain} and \eqref{eq:closedForm}. Once $\FisherInformationMatrix$ is calculated, \ac{CRLB} for the receiver location in 3D geometry is obtained as
\def\RMSE[#1]{{\rm RMSE}(#1)}
\begin{align}
\RMSE[x,y,z] \ge\sqrt{{\rm tr}(\inverseFisherInformationMatrix)}~,
\label{eq:CRBL_RSSI}
\end{align}
where $\RMSE[x,y,z]$ is the \ac{RMSE} of the estimator given in \eqref{eq:NLS}. It is worth noting that the analytic expression of $\jacobianMatrix$ allows arbitrary scenarios in 3D coordinates, which generalizes the theoretical  results provided in \cite{Zhang_2014} where the angle between detector plane and floor plane is fixed to $0$ degrees.

\section{Numerical Results}
\label{sec:numericalResults}

\begin{figure}[!t]
\centering
{\includegraphics[width =3.5in]{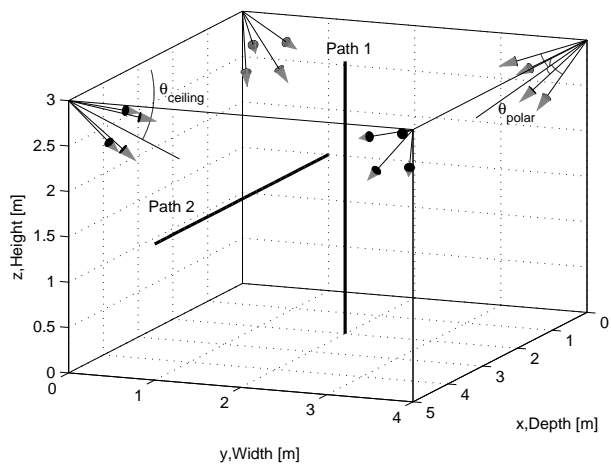}}
\caption{Simulation setup.}
\label{fig:simSetup}
\end{figure}

\def\widthFig{3.5in}
\begin{figure*}[t]
\centering
\subfloat[Path 1,  $\mode=10$.]{\includegraphics[width=\widthFig]{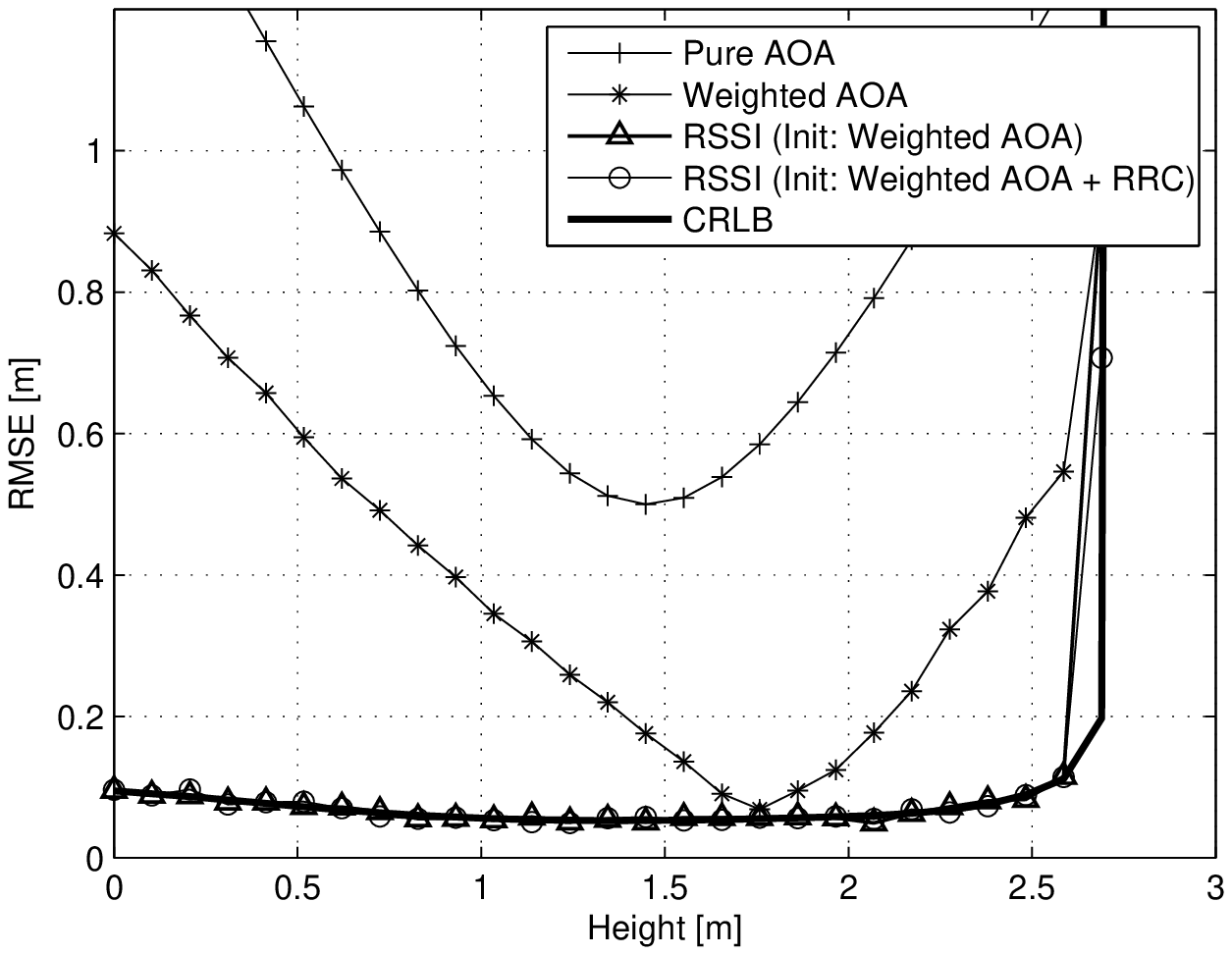}
\label{fig:path1n10}}
\subfloat[Path 1, $\mode=30$.]{\includegraphics[width=\widthFig]{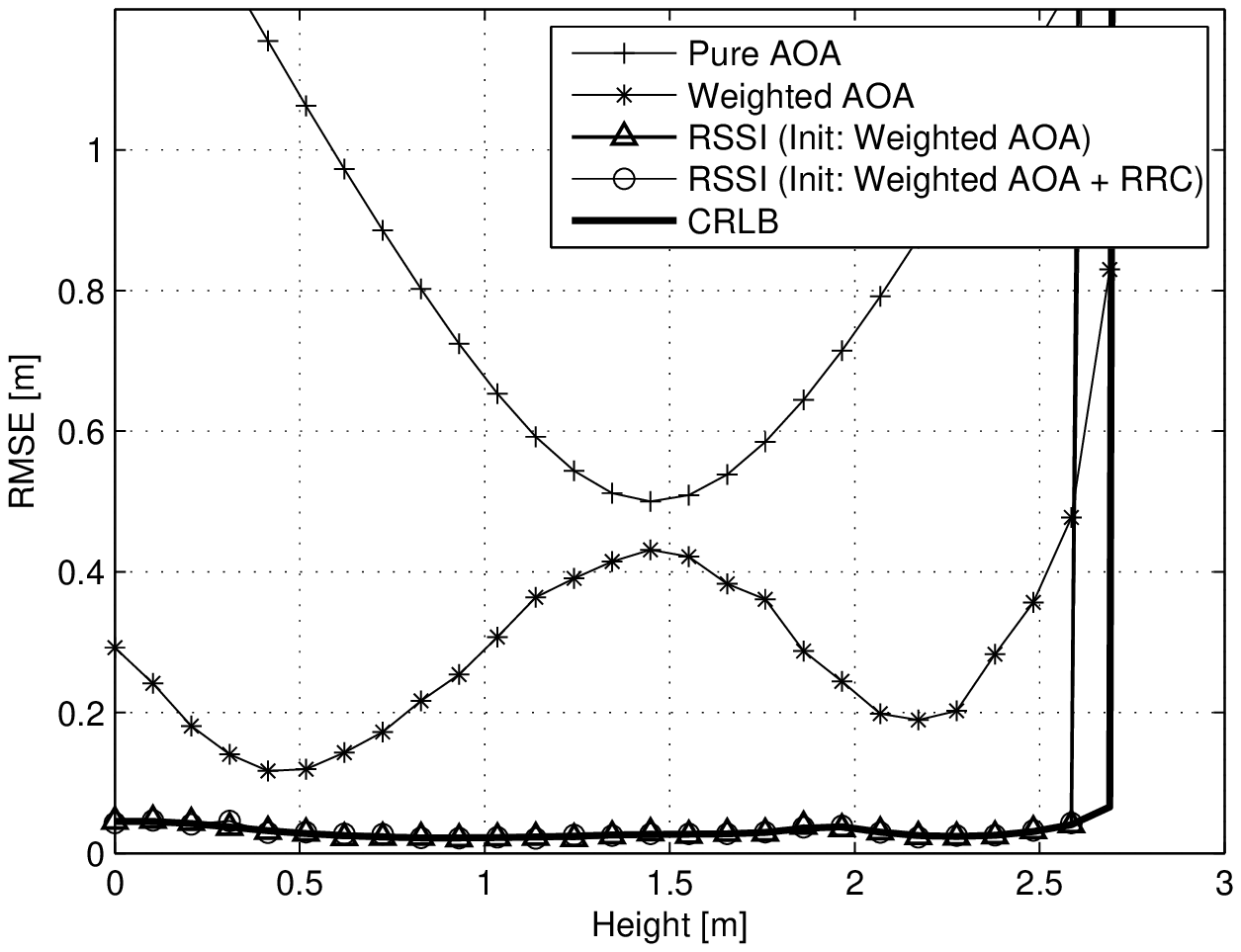}
\label{fig:path1n30}}\\
\subfloat[Path 2,  $\mode=10$.]{\includegraphics[width=\widthFig]{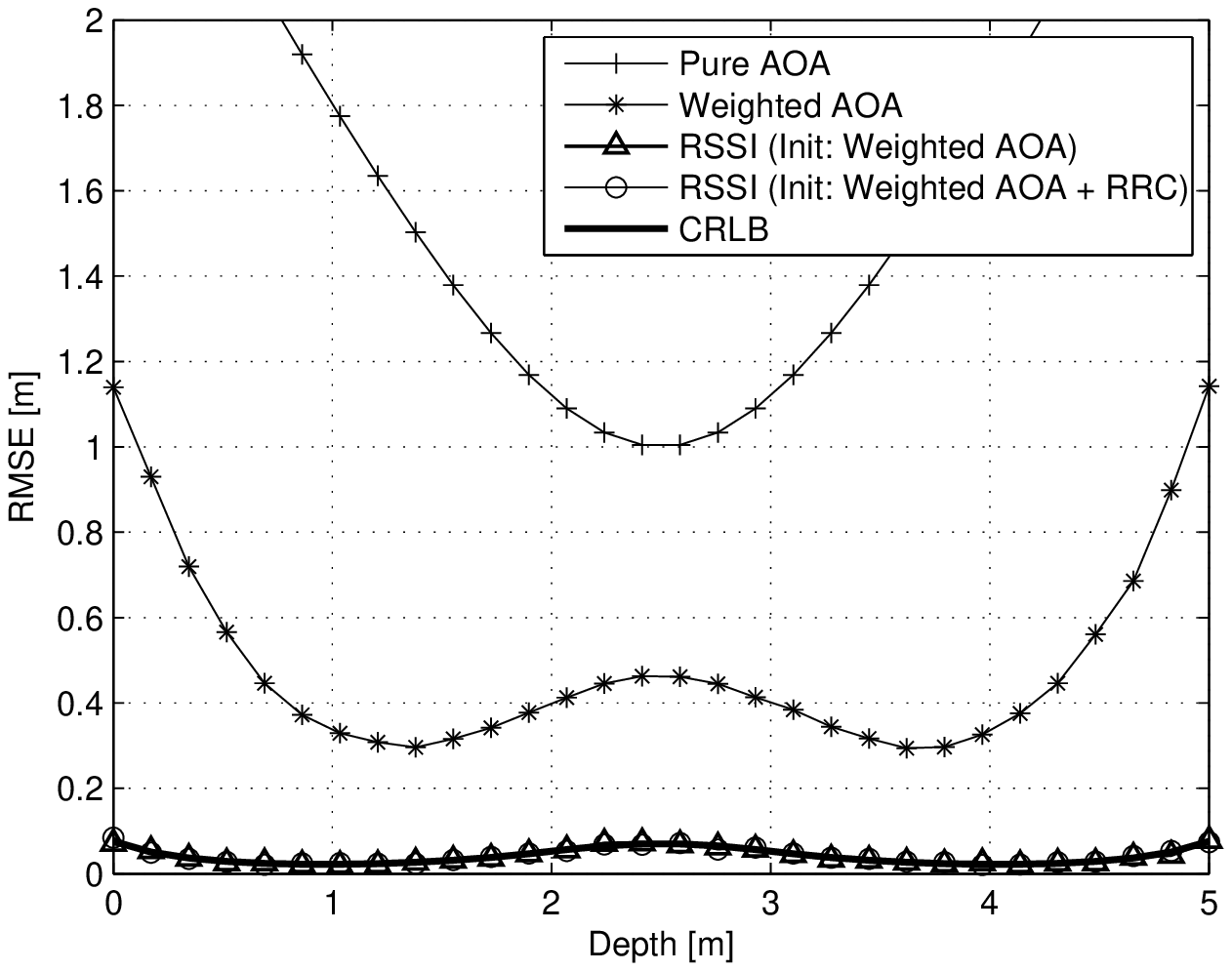}
\label{fig:path2n10}}
\subfloat[Path 2, $\mode=30$.]{\includegraphics[width=\widthFig]{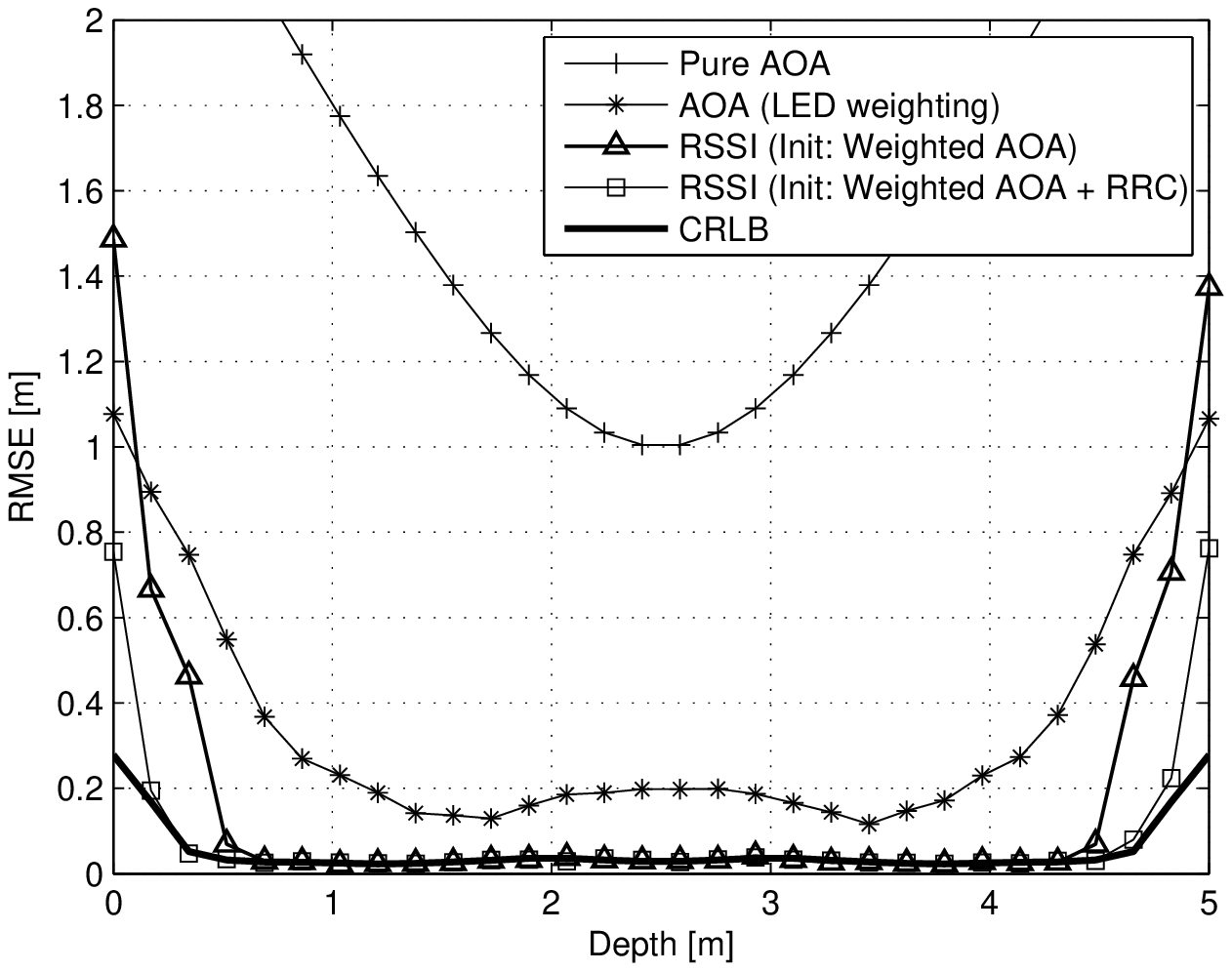}
\label{fig:path2n30}}
\caption{Positioning accuracy on Path 1 and Path 2 ($\polarAngle=20^\circ$, $\ceilingAngle=30^\circ$).}
\label{fig:acc}
\end{figure*}

In this section, we evaluate the performance of \ac{AOA} and \ac{RSS} based  methods and impact of physical characteristics of \ac{LED} transmitters on the estimators through computer
simulations. For simulation tractability, we consider an empty room where its depth, width, and height are set to $5$, $4$, and $3$ meters, respectively. We deploy $4$ \acp{VAP} which are located at the corners of the room as shown in \figurename~\ref{fig:simSetup}. The angle between the direction of \ac{VAP} and ceiling is denoted by $\ceilingAngle$. Each \ac{VAP} has $4$ \ac{LED} transmitters where the angle between the direction of \ac{VAP} and the direction of \ac{LED} transmitters is assumed to be identical and parameterized as $\polarAngle$.  For the \ac{VLC} receiver, $\FOV$ and $\Ar$ are set to $85$ degrees and $1$~cm$^2$, respectively. The direction of the \ac{VLC} receiver is assumed to be $\directionDetector=[0,0,1]^{\rm T}$.  Finally, based on the noise model given in \cite{Zhang_2014} and the references therein, we set $\noiseVariance$ to $1$e$-13$~A$^2$.

\subsection{Performance of AOA and RSS Based Location Estimators}

In order to  evaluate the estimators discussed in Section \ref{sec:AOA} and Section \ref{sec:RSSI} numerically, we consider two different paths, i.e., Path 1 and Path 2, as shown in \figurename~\ref{fig:simSetup}. Path 1 considers a \ac{VLC} receiver  where its localization on x-axis and y-axis are fixed to $2$~m. On the other hand, Path 2 follows a horizontal path where the exact position of the \ac{VLC} receiver on y-axis and z-axis are set to $1$~m and $1.5$~m, respectively. We then sweep the \ac{VLC} receiver's position on z-axis for Path 1 and x-axis for Path 2 and calculate \ac{RMSE} of each estimator after $50$ realizations. We perform the simulations for $\mode=10$ and $\mode =30$ in \eqref{fig:systemModel} when $\ceilingAngle = 30^\circ$ and $\polarAngle = 20^\circ$.

As it can be seen from \figurename~\ref{fig:acc}, \ac{LED} mode, i.e., $\mode$, does not affect the performance of pure \ac{AOA} method based on \eqref{eq:CBsolutionLS} for both Path 1 and Path 2. This result is expected since the \ac{LED} transmitters are assumed to be identical. Therefore \ac{VLC} receiver selects the same \ac{LED} transmitters as anchors regardless of $\mode$, which yields the same estimation results for pure \ac{AOA} method. However, \ac{LED} directivity, characterized by $\mode$, significantly affects the performance of \ac{AOA}-based estimation,  when \ac{RSS} information is taken into account. For weighted \ac{AOA} method based on  \eqref{eq:CBsolution}, increasing $\mode$ provides better estimation results on both Path 1 and Path 2. This is due to the fact that \ac{RSS} information becomes a dominant factor when \ac{VLC} receiver is close to the line pointed by the orientation vectors of \ac{LED} transmitters. In other words, nodes with higher $\mode$ have more reliable \ac{AOA} estimates for a given \ac{RSS} value, which is captured by the proposed weighted \ac{LS} estimator. 

For \ac{RSS}-based localization, we consider two different initializations for the Newton-Raphson method ($\stepSize =0.2$) given in \eqref{eq:update}. We first run the simulations when weighted \ac{AOA} results are considered as the initialization points. In this case, the search algorithm may happen to converge to local optima, which causes significant positioning errors. For example, the estimator performance degrades drastically  for Path 2 when the \ac{VLC} receiver is close to the side walls of the room, as given in  \figurename~\ref{fig:acc}\subref{fig:path1n30}. On the other hand, when \ac{RRC} algorithm ($\numberOfRandomPoints = 500$, $\numberOfselectPoints=100$, and $\numberOfClusters = 4$) is applied for finding better initial points, \ac{RSS} based localization is more likely to find the global optimum and attains the \ac{CRLB} in most of the positions.

When the position of \ac{VLC} receiver is close to the ceiling,  all methods that use \ac{RSS} fail as shown in \figurename~\ref{fig:acc}\subref{fig:path1n10} and  \figurename~\ref{fig:acc}\subref{fig:path1n30}. This is due to the large angle between receiver orientation vector and  the incidence vector  which reduces the effective area of \ac{PD}. In particular, after certain height, \ac{FOV} of \ac{VLC} receiver does not allow the \ac{PD} to capture any signal, which explains why \ac{CRLB} goes to infinity.

\subsection{Number Of Clusters for RRC Algorithm}
In this subsection, we evaluate the impact of number cluster of \ac{RRC} algorithm, based on the configuration given in \figurename~\ref{fig:simSetup}.  As a pessimistic scenario, we consider a \ac{VLC} receiver located at the origin of the coordinate system. We then provide the probability of convergence to the location of \ac{VLC} receiver  for $\numberOfClusters=\{0,1,2,3,4\}$ when $\mode=10$ and $\mode=30$. For each scenario, we also include the result of weighted \ac{AOA} to the initial points obtained via \ac{RRC} algorithm. In order to have a better understanding on the convergence, we evaluate the impact of number clusters in the noiseless case. We assume that a successful converge is considered when $\norm[\locationDetectorEstimate - \locationDetector]_2^2$ is less than $1$e$-2$. 

\begin{figure}[t]
\centering
{\includegraphics[width=\widthFig]{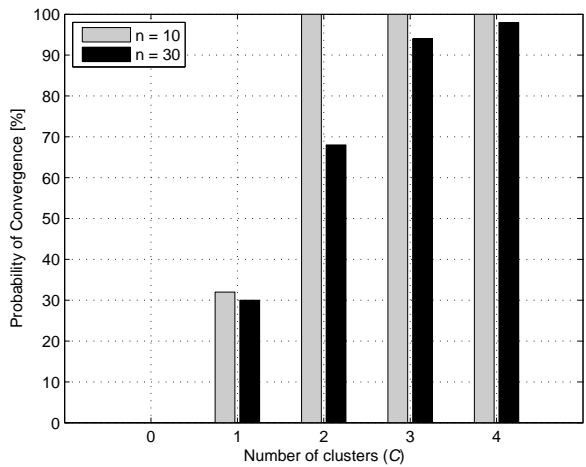}}
\caption{Probability of convergence for RRC algorithm.}
\label{fig:convergence}
\end{figure}

As shown in \figurename~\ref{fig:convergence}, increasing number of clusters yields better probability of converge. However, different \ac{LED} modes require different number of clusters. For example, while \ac{RRC} with $2$ clusters is sufficient to attain the global optimum with very high probability when $\mode=10$, \ac{RRC} with $4$ cluster, i.e., 4 extra initial points beside weighted \ac{AOA} result, yields high probability of convergence to global optimum when $\mode=30$.

\subsection{Configuration of VLC Access Points}

\label{subsec:deployment}
\begin{figure}[t]
\centering
\subfloat[$\mode=10$.]{\includegraphics[width=\widthFig]{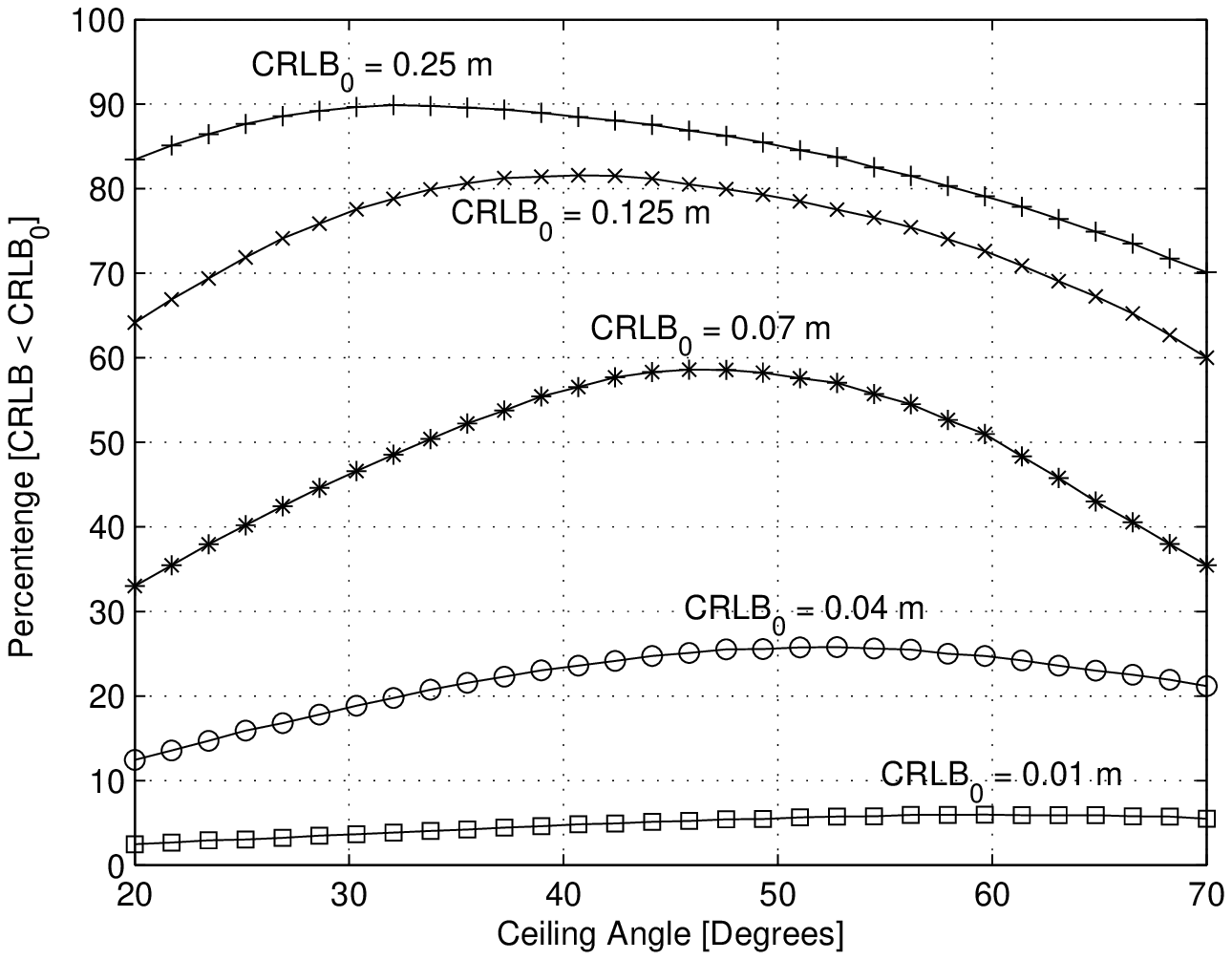}
\label{fig:n10}}\\
\subfloat[ $\mode=30$.]{\includegraphics[width=\widthFig]{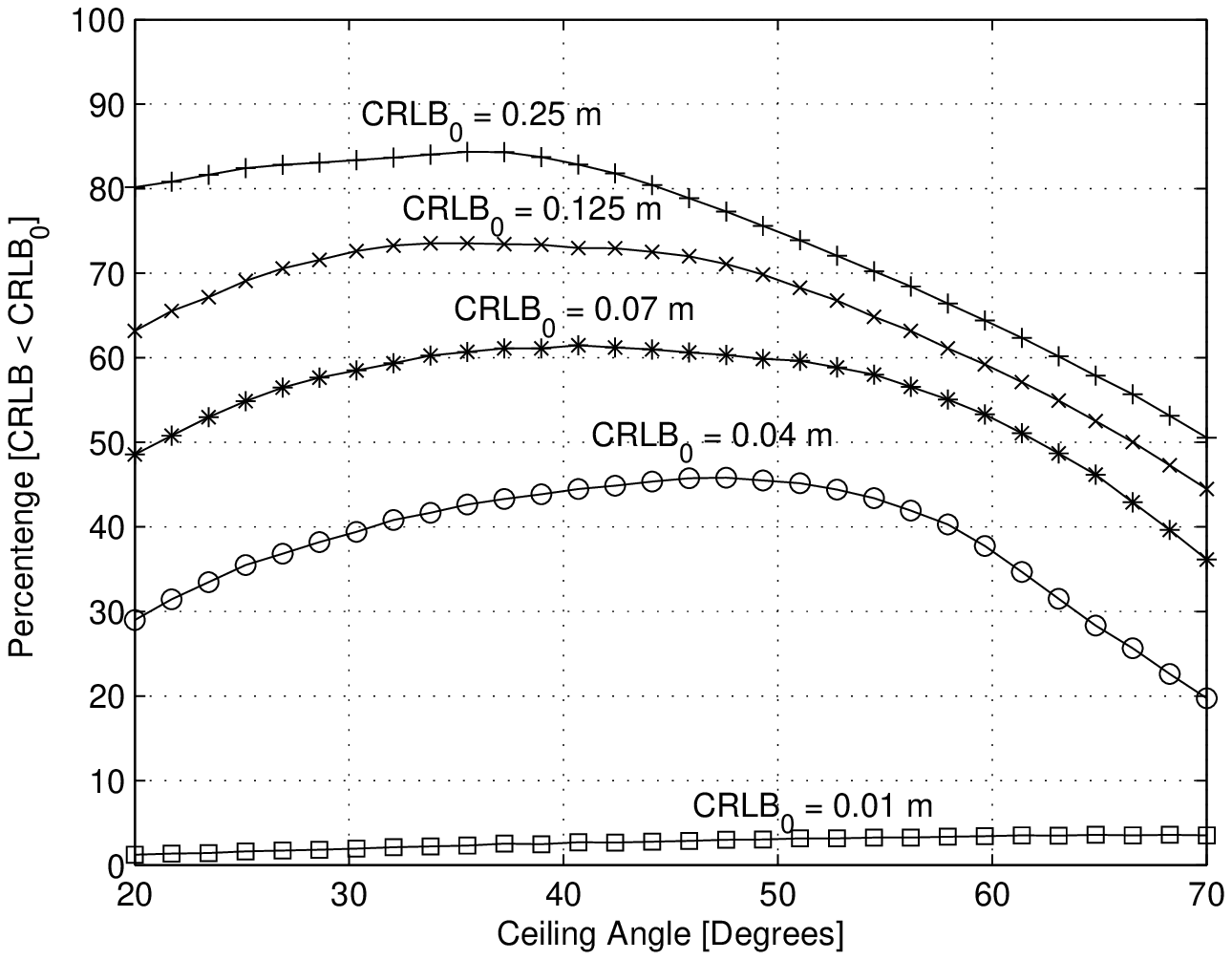}
\label{fig:n30}}
\caption{Probability of accuracy for a given $\ceilingAngle$ ($\polarAngle=20^\circ$).}
\label{fig:cov}
\end{figure}

In this subsection, we investigate the impact of deployment of \ac{VAP} on positioning accuracy. We consider the simulation setup illustrated in \figurename~\ref{fig:simSetup} and choose $\ceilingAngle$ as a design parameter. We sweep $\ceilingAngle$ between $20^\circ$ and $70^\circ$ and calculate \acp{CRLB} for the positions of the \ac{VLC} receiver based on 3D grid in the whole room. We then calculate the probabilities where \ac{CRLB} is less than or equal to certain values, i.e.,  ${\rm CRLB}_0=\{0.01, 0.04, 0.07, 0.125,0.25\}$ in meters when $\mode=10$ and $\mode =30$. In the analysis, we fix $\polarAngle$ to $20^\circ$. 

As it can be seen from both \figurename~\ref{fig:cov}\subref{fig:n10} and  \figurename~\ref{fig:cov}\subref{fig:n30}, the optimum $\ceilingAngle$ that gives the highest probability of accuracy in the room varies depending on $\rm CRLB_0$ and \ac{LED} mode $\mode$. For example, the highest probability of accuracy  is obtained at $\ceilingAngle =60^\circ$ when $\rm CRLB_0 =0.01$~m  as given in \figurename~\ref{fig:cov}\subref{fig:n10}. On the other hand, the highest probability is achieved at $\ceilingAngle=30^\circ$  for the same \ac{LED} mode when $\rm CRLB_0 =0.25$~m. Similarly, when $\mode$ is set to $30$, the choice of  $\ceilingAngle=37^\circ$ yields the highest probability of accuracy for the same $\rm CRLB_0$ as given in \figurename~\ref{fig:cov}\subref{fig:n30}.

It is also possible to infer how \ac{LED} directivity affects the positioning accuracy overall in the room from \figurename~\ref{fig:cov}. As it can be seen from \figurename~\ref{fig:cov}\subref{fig:n10}, $\%90$ of the room is covered when $\rm CRLB_0 =0.25$~m and $\mode=10$. However, increasing \ac{LED} mode $\mode$ to $30$ reduces the probability of accuracy to $\%85$ of the room as given in \figurename~\ref{fig:cov}\subref{fig:n30}. On the other hand, higher \ac{LED} mode increases the probability of highly accurate results. For example, when ${\rm CRLB}_0$ is fixed to $0.04$~m, $\%45$ of the room is covered for $\mode=30$ while it reduces to $\%27$ of the room for $\mode=10$.

\subsection{Orientations of LED Transmitters}
\begin{figure}[t]
\centering
\subfloat[$\mode=10$.]{\includegraphics[width=\widthFig]{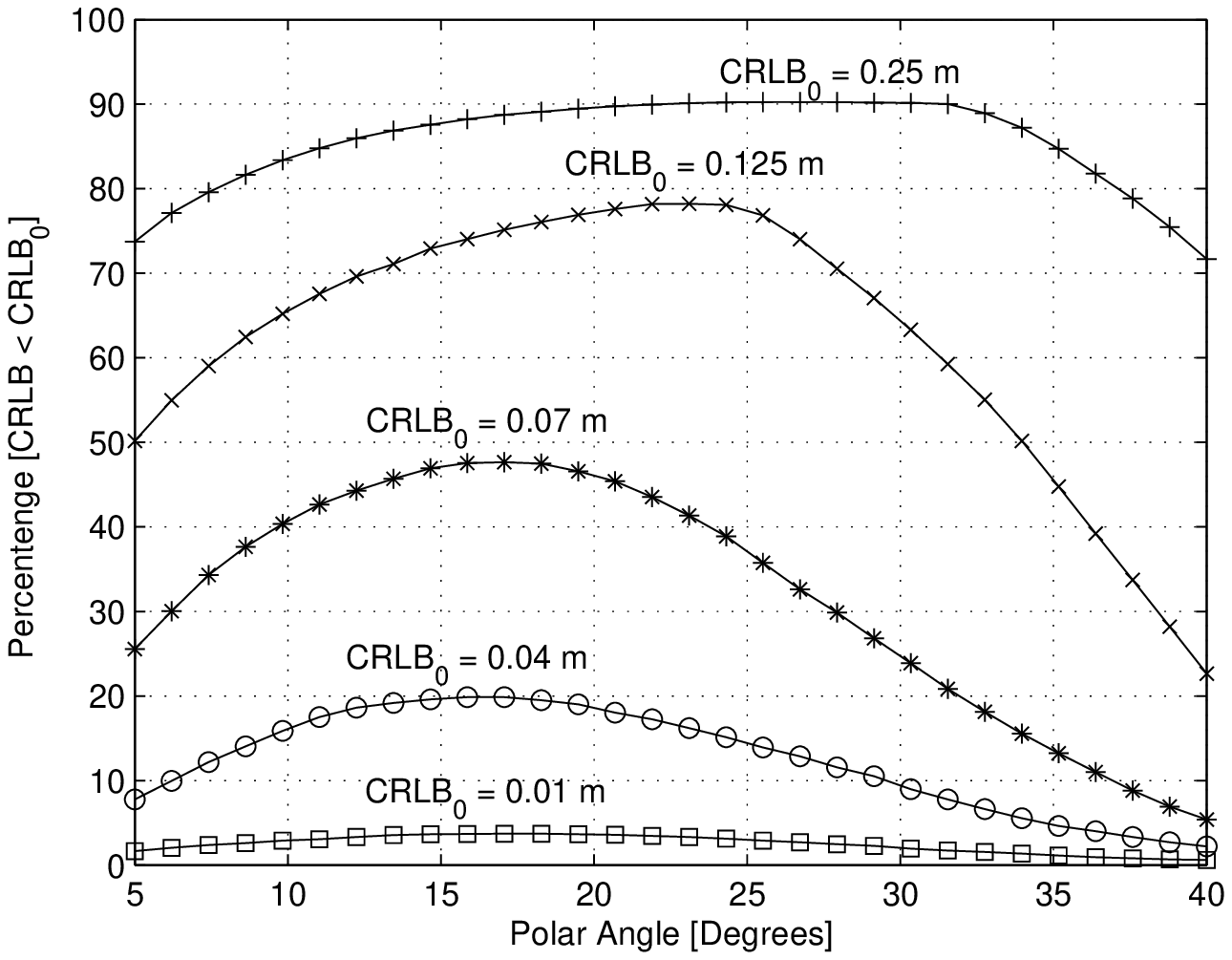}
\label{fig:n10_polar}}\\
\subfloat[ $\mode=30$.]{\includegraphics[width=\widthFig]{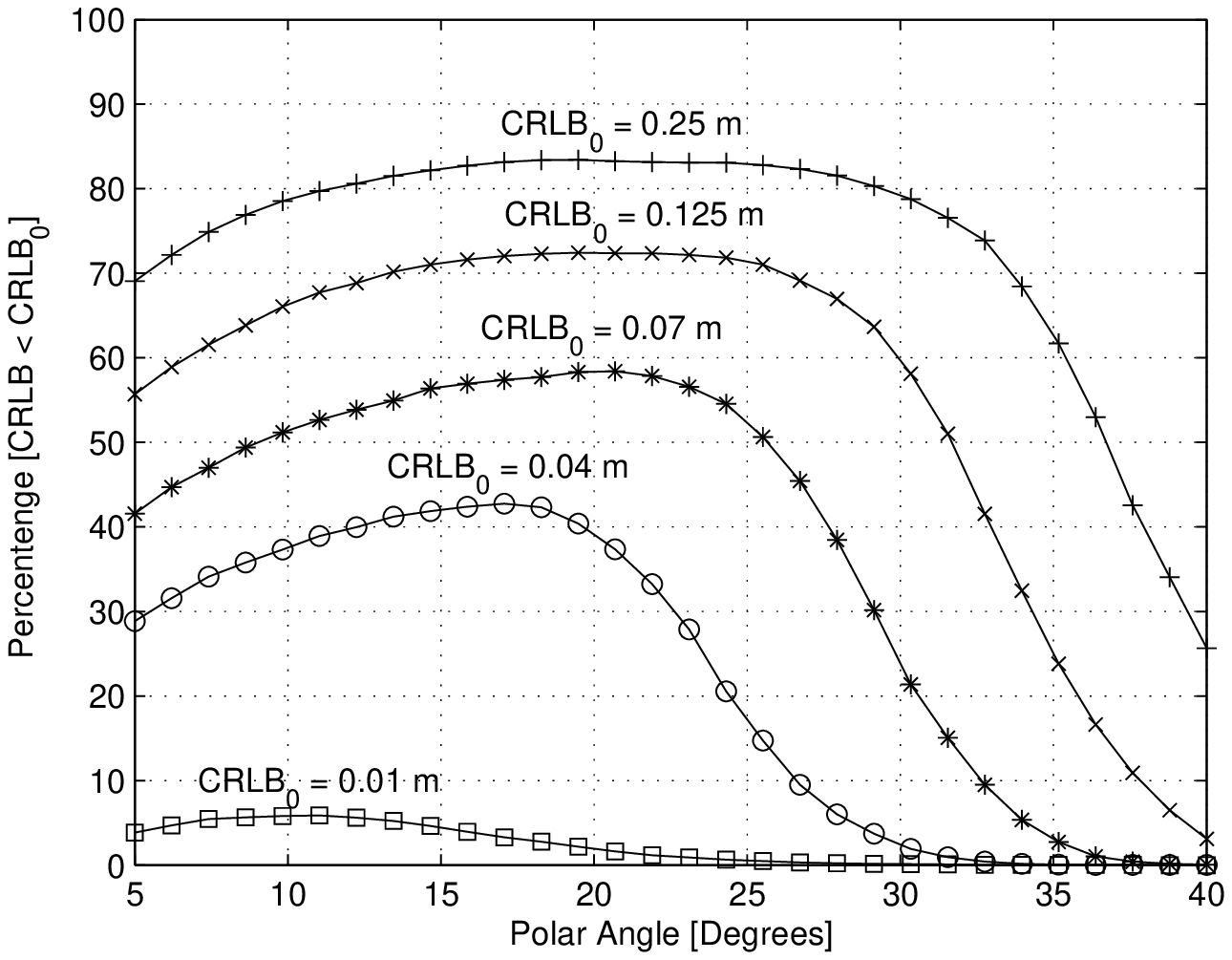}
\label{fig:n30_polar}}
\caption{Probability of accuracy for a given $\polarAngle$ ($\ceilingAngle=30^\circ$).}
\label{fig:cov_polar}
\end{figure}

In this subsection, we discuss the impact of the orientations of \ac{LED} transmitters on positioning accuracy. Considering the setup illustrated in \figurename~\ref{fig:simSetup}, we sweep $\polarAngle$ between $5^\circ$ and $40^\circ$. Similar to the analysis given in Section \ref{subsec:deployment}, we obtain the probabilities that \ac{CRLB} is less than or equal to certain values when $\mode=10$ and $\mode =30$. For this analysis, we fix $\ceilingAngle$ to $30^\circ$.

As shown in \figurename~\ref{fig:cov_polar}, the optimum $\polarAngle$ depends on the \ac{LED} mode and selected $\rm CRLB_0$. For instance, when $\mode = 10$ and $\rm CRLB_0=0.01$~m, the optimum $\polarAngle$ is $17.5^\circ$ as given in \figurename~\ref{fig:cov_polar}\subref{fig:n10_polar}. However, as shown in \figurename~\ref{fig:cov_polar}\subref{fig:n30_polar}, the optimum $\polarAngle$ is obtained as $10^\circ$ when  $\mode = 30$. Similarly, when  $\rm CRLB_0$ is set to $0.25$~m, $27^\circ$ and $19^\circ$ yield to the highest probability of accuracies in the room when  $\mode = 10$ and  $\mode = 30$, respectively. Nevertheless, in general, the optimum $\polarAngle$ increases for higher ${\rm CRLB}_0$. This is due to the fact that larger $\polarAngle$ yields more physical separation between \ac{LED} transmitters for each \ac{VAP} and the  intensity distribution becomes more homogeneous in the room. Therefore while narrow $\polarAngle$ yields more accurate results at specific points of the room, larger $\polarAngle$ provides less accurate but better probability of accuracy for the setup given in \figurename~\ref{fig:simSetup}. On the other hand, increasing $\polarAngle$ above the optimal point deteriorates the positioning accuracy as the orientations of \ac{LED} transmitters start to be parallel to the walls of the room. In these cases, the center of the room does not receive sufficient amount of energy, which causes more positioning error.

\section{Conclusion}
\label{sec:conclusions}
In this study, we discuss \ac{AOA}-based localization and \ac{RSS}-based localization methods by considering hybrid utilization of \ac{AOA} and \ac{RSS} information in the location estimation. We show that \ac{AOA}-based localization can be solved with an \ac{LS} estimator. Yet its estimation results may be highly inaccurate, depending on the location of \ac{VLC} receiver. On the other hand, when \ac{RSS} information is utilized to weight \ac{LED} transmitters in the optimization, the positioning accuracy increases significantly. For the scenario investigated in this study, i.e. a room with the dimensions of $3\times 5\times 4$ meters and $4$ \acp{VAP}, the localization error is less than $1$ meter. On the other hand, \ac{RSS}-based localization method, which exploits the Lambertian patterns of \acp{LED}, offers high positioning accuracy at the expense of a system of nonlinear equations and a non-convex objective function. In order to solve the system of nonlinear equations, we derive an analytical learning rule based on the Newton-Raphson method. Since the learning rule is analytical, it eliminates the calculation of Jacobian matrix numerically. For the hybrid utilization, the learning rule is initialized with the result of \ac{AOA}-based method to increase the likelihood of converging to the global optimum, which the positioning accuracy improves up to $10$ cm. In addition to employing the results of \ac{AOA}-based method, we also develop a heuristic search method, \ac{RRC} algorithm, to identify extra initial points which could lead to find the global optimum.

In this investigation, we also discuss the impact of the orientations of \ac{LED} transmitters and the physical characteristics of \acp{LED} on the localization performance probabilistically. For this purpose, we utilize the \ac{CRLB} that is derived for an arbitrary configuration in 3D geometry in this study.  According to our analyses, when the illumination is homogeneous in 3D geometry, the positioning error becomes relatively high but mostly in attainable levels. On the other hand, when the illumination is not homogeneous due to the orientations of \acp{LED} or using highly directive \acp{LED},  the positioning accuracy is improved significantly at the locations where the energy is high; this also degrades the positioning accuracy at the same time in the locations where the energy, i.e., illumination, is low.

%
\bibliographystyle{IEEEtran}
\bibliography{journalCRLB}

\end{document}